\documentclass[a4paper,fleqn,usenatbib]{mnras}
\pdfoutput=1

\usepackage[T1]{fontenc}
\usepackage{ae,aecompl}

\usepackage{graphicx}	
\usepackage{amsmath}	
\usepackage{amssymb}	
\usepackage{bm}
\usepackage{multirow}
\usepackage{comment}

\title[Accretion rate in circumbinary discs]{Suppression of the accretion rate in thin discs around binary black holes}

\author[E. Ragusa, G. Lodato, D. Price]{
Enrico Ragusa,$^{1}$\thanks{E-mail: enrico.ragusa@unimi.it}
Giuseppe Lodato,$^{1}$\thanks{E-mail: giuseppe.lodato@unimi.it}
Daniel J. Price,$^{2}$\thanks{E-mail: daniel.price@monash.edu}
\\
$^{1}$ Dipartimento di Fisica, Universit\`a degli studi di Milano, Via Celoria 16, Milano, I-20133, Italy\\
$^{2}$Monash Centre for Astrophysics (MoCA), School of Physics and Astronomy, Monash University, Vic 3800, Australia
}

\date{Accepted XXX. Received YYY; in original form ZZZ}

\pubyear{2016}

\begin{document}
\label{firstpage}
\pagerange{\pageref{firstpage}--\pageref{lastpage}}
\maketitle

\begin{abstract}

We present three-dimensional Smoothed Particle Hydrodynamics (SPH) simulations investigating the dependence of the accretion rate on the disc thickness around an equal-mass, circular black hole binary system. 
We find that for thick/hot discs, with $H/R\gtrsim 0.1$, the binary torque does not prevent the gas from penetrating the cavity formed in the disc by the binary (in line with previous investigations). The situation drastically changes for thinner discs, in this case the mass accretion rate is suppressed, such that only a fraction (linearly dependent on $H/R$) of the available gas is able to flow within the cavity and accrete on to the binary.
Extrapolating this result to the cold and thin accretion discs expected around supermassive black hole binary systems implies that this kind of systems accretes less material than predicted so far, with consequences not only for the electromagnetic and gravitational waves emissions during the late inspiral phase but also for the recoil speed of the black hole formed after binary coalescence, thus influencing also the evolutionary path both of the binary and of the host galaxy. Our results, being scale-free, are also applicable to equal mass, circular binaries of stellar mass black holes, such as the progenitor of the recently discovered gravitational wave source GW150914.

\end{abstract}




\section{Introduction}

The $\Lambda$CDM cosmological model predicts the ubiquitous formation of supermassive black hole binary systems. According to this model, galaxies in the near universe were assembled via the hierarchical merger of smaller structures at high redshift. Most galaxies host in their central regions a supermassive black hole \citep{2013ARA&A..51..511K}.
Hence, it is natural to assume that after a galaxy merger two supermassive black holes coexist in the same galaxy. The observation of dual AGNs with separations of the order of $1-10\, {\rm kpc}$ (\citealp{2015ApJ...806..219C}, and references therein) appears to support this hypothesis; however the lack of observations of binary systems at sub-pc separations (apart from few candidates, \citealp{2015ASSP...40..103B}) suggests a rapid orbital decay of the binary, until the supermassive black holes coalesce on a reasonably short time-scale \citep{1980Natur.287..307B}. This explanation has been recently strengthened by the constraints placed on the GWB (gravitational wave background) through the PTA (Pulsar Timing Array) technique \citep{2015Sci...349.1522S, 2016ApJ...821...13A, 2015MNRAS.453.2576L}, suggesting that the binary shrinking could be even faster than so far predicted.

When the separation is of the order of $\sim 10 \, {\rm pc}$ the black holes become gravitationally bound in an ``hard'' binary configuration \citep{2002MNRAS.331..935Y}.
For separations $\lesssim 10\, {\rm pc}$, the interaction with the surrounding gaseous environment\footnote{A gaseous disc forms in the centre of the new galaxy as a consequence of angular momentum conservation \citep{2002MNRAS.333..481B}, if the parent galaxies are gas rich.} is able to further reduce the binary separation by gas friction \citep{2005ApJ...630..152E, 2007Sci...316.1874M, 2007MNRAS.379..956D, 2009MNRAS.396.1640D} and mechanisms similar to those predicted for planetary migration in young solar systems \citep{2008ApJ...672...83M, 2009MNRAS.393.1423C,2009MNRAS.398.1392L}, such as Type II migration \citep{1986ApJ...309..846L,1997Icar..126..261W,2002ApJ...565.1257T}; this part of the binary evolution is the so called ``final parsec problem'' \citep{2009MNRAS.398.1392L}.
The emission of gravitational waves becomes the dominant migration mechanism for separations of $\lesssim 0.001 \,pc$, driving the binary to coalescence.  

During the gas driven migration phase, the accretion rate on to the black holes is the primary determinant of the evolution of the binary properties. It influences the black hole spin alignment process \citep{2007ApJ...661L.147B,2013ApJ...774...43M,2013MNRAS.429L..30L,2015MNRAS.451.3941G}, and thus determines the gravitational wave frequency pattern and the recoil velocity of the black hole formed after the coalescence \citep{2016MNRAS.456..961B}. 

An accurate determination of the accretion rate on to supermassive black hole binaries at sub-pc scales is needed to infer their luminosities and in general their electromagnetic output and is thus key to interpreting observations of candidate binaries \citep{2015Natur.525..351D} and to discriminate between various interpretations (e.g., recoiling versus binary black holes). 

The issue of the mass flow within the cavity formed by a binary in its surrounding disc has been mostly studied in the protostellar case (e.g. \citealp{1994ApJ...421..651A, 1996ApJ...467L..77A}) where the disc aspect ratio $H/R$ is relatively large. In recent years, the interest has moved to the supermassive black hole binary case. However, mostly for numerical reasons, simulations of these systems have used $H/R\sim 0.1$, appropriate for protostellar binaries, but two orders of magnitude larger than the values expected for a disc surrounding a supermassive black hole binary. These studies all conclude that the binary potential does not prevent the gas from flowing within the cavity so that the accretion rate on to the binary, $\dot M_{\rm bin}$, is comparable to the equivalent rate for a single object, $\dot M_0$ \citep{2012A&A...545A.127R, 2012ApJ...749..118S, 2013MNRAS.436.2997D, 2014ApJ...783..134F, 2015ApJ...807..131S}. Here we investigate how this conclusion is modified when one adopts more realistic values for the disc temperature and aspect ratio. 

The paper is organized as follows: in Section \ref{accrinbinsys} we summarize the known results for the dynamics of gas orbiting a binary system. Section \ref{numsim} describes the numerical method used in our simulations and the initial conditions. In Section \ref{resu} we first reproduce the results of earlier works using a large $H/R$ and then show how the results change when reducing $H/R$. We discuss our results and draw conclusions in Section \ref{concl}.

\section{Accretion dynamics in binary systems}\label{accrinbinsys}

Adapting classical accretion disc theory, developed for a single central massive object, to binary systems is far from trivial. The tidal torques produced by the disc-satellite interaction \citep{1979MNRAS.186..799L,1980ApJ...241..425G}, strongly perturb the disc structure \citep{1994ApJ...421..651A} repelling the gas from the corotation region and clearing an annular gap across the orbit of the secondary object.

For sufficiently high mass ratios, the gap becomes so wide that the binary resides in a depleted cavity surrounded by a circumbinary disc. The effects of these strong tidal forces on the accretion are uncertain. Do they act as a dam or is the gas able to leak into the cavity from the circumbinary disc? 

An analytical treatment of the gas dynamics in binary systems has been attempted \citep{1991MNRAS.248..754P, 2010PhRvD..82l3011L}, predicting a suppression of the accretion rate for high values of the binary mass ratio $q=M_2/M_1>0.01$, where $M_1$ and $M_2$ are the primary and the secondary mass, respectively.
However these models assume, being 1D, axial symmetry, which is not the case for a rotating binary potential.

Starting from the mid 1990s, the problem of accretion in binary systems has been treated numerically for both binary-star/planet-star systems \citep{1996ApJ...467L..77A,1997MNRAS.285...33B,2002A&A...387..550G,2005ApJ...623..922O,2010ApJ...708..485H,2011MNRAS.413.2679D,2015MNRAS.448.3545D} and supermassive black hole binaries \citep{2007PASJ...59..427H, 2008ApJ...672...83M, 2009MNRAS.393.1423C, 2012A&A...545A.127R, 2012ApJ...749..118S, 2013MNRAS.436.2997D, 2014ApJ...783..134F, 2015ApJ...807..131S}.

These simulations showed that for aspect ratios $H/R\sim 0.1$ one or two streams of material --- the number depending on the mass ratio $q$ --- flow through the edge of the cavity. These streams connect the binary with the edge of the cavity; their formation is periodic with characteristic frequencies, $\omega_{\rm stream}/\Omega_{\rm bin}\simeq1-2$, where $\Omega_{\rm bin}$ is the binary frequency. 

Various authors \citep{2008ApJ...672...83M,2013MNRAS.436.2997D, 2014ApJ...783..134F} noted the development of an eccentric cavity associated with an overdense lump of material at the edge of the cavity. This lump orbits around the binary on an elliptic Keplerian orbit. When it reaches the pericentre of its orbit it causes a boost in the accretion, therefore adding also a lower characteristic frequency $\omega_{\rm lump}/\Omega_{\rm bin}=2/9$. Recently \citet{2014ApJ...783..134F} showed that the accumulation of gas around each object in circum-individual ``mini-discs'' acts as a buffer for accretion, smoothing the faster periodicity and increasing the power of the $\omega_{\rm lump}/\Omega_{\rm bin}=2/9$ one, with important consequences for observations.

The accretion rate on to the binary $\dot M_{\rm bin}$ has been found to be comparable to that predicted by classic disc accretion theory $\dot M_0$ in absence of the binary companion \citep{2012ApJ...749..118S, 2012A&A...545A.127R, 2013MNRAS.436.2997D, 2014ApJ...783..134F, 2015ApJ...807..131S}, giving a normalized accretion rate $1<\dot M_{\rm bin}/\dot M_0\lesssim 1.6$, implying, de facto, that no significant suppression of the accretion occurs due to the presence of the binary. However, as we discuss, it is not surprising that lower values of $\dot M_{\rm bin}/\dot M_0$ are associated with lower viscous torques \citep{2008ApJ...672...83M,2013MNRAS.436.2997D}, or in general with less effective angular momentum transfer mechanisms in the disc.

\citet{2012ApJ...749..118S} and \citet{2015ApJ...807..131S} investigated this issue, trying to constrain the uncertainty on the viscous $\alpha$-parameter (that in the literature typically spans the values $0.01\leq\alpha\leq 0.1$) by performing MHD simulations, in order to provide a self-consistent physical mechanism for the angular momentum transport through the disc. They found that the angular momentum transport operated by the turbulent motion of the gas fed by MRI, is consistent with a corresponding viscous $\alpha=0.1$ for an equal mass, circular binary system and a globally isothermal disc with $c_{\rm s}=0.1\Omega_{\rm bin}a$, where $a$ is the binary separation. 

Regarding the differential accretion rate, i.e. how much material is accreted by each object, recent results claim that, for a fixed mass ratio $q$ ranging between $0.1\leq q<1$, the ratio $\dot M_2/(\dot M_1 +\dot M_2)$ gets higher as the disc gets colder \citep{2015MNRAS.452.3085Y, 2015MNRAS.447.2907Y}, where $\dot M_1$ and $\dot M_2$ are the accretion rate on the primary and secondary object, respectively. Accretion occurs mostly on the secondary object since it is closer to the cavity wall than the primary, but for high gas temperatures pressure allows the material to cross the L1 Lagrange point, thus accreting on the primary and reducing differential accretion\footnote{This result resolved a long standing issue regarding the discrepancies in the literature between some works \citep{1997MNRAS.285...33B, 2005ApJ...623..922O,2010ApJ...708..485H}, attributing the qualitative and quantitative differences in the results to the different gas temperatures used in the simulations.}. For $q<0.1$ the accretion returns progressively to occur mostly on the primary object \citep{2014ApJ...783..134F}.
 
We consider in this paper coplanar prograde discs. Misaligned or even retrograde discs \citep{2012MNRAS.423.2597N, 2013MNRAS.434.1946N, 2015ApJ...800...96L, 2014MNRAS.445.2285D, 2015MNRAS.449...65A, 2015MNRAS.448.3472N, 2016MNRAS.455.1989G} have also been considered, demonstrating the importance of disc tearing and retrograde accretion in the evolution of such systems.

\subsection{The problem of the accretion in supermassive black hole binary systems}\label{probaccbin}
  
Analytical models of binary systems with discs assume that the presence of a binary companion gives rise to tidal torques acting on the disc.
Using the \textit{impulse approximation} \citep{1979MNRAS.186..799L}, developed under the assumption $q\ll1$, the tidal torque density exerted by a satellite on the circumbinary disc may be approximated by
\begin{equation}
\frac{\partial T_{\rm tid}}{\partial R}=\pi f q^2 \Omega^2 R^3\Sigma \left(\frac{a}{\Delta}\right)^4,\label{ttid}
\end{equation}
where  $\Omega$ is the Keplerian angular frequency, $a$ is the binary separation, $R$ is the cylindrical radius, $\Sigma$ the surface density, $f$ is a dimensionless normalization factor and $\Delta=\max[R-a,H,R_{\rm Hill}]$, where $H$ is the disc height and $R_{\rm H}$ is the Hill radius \citep{1995MNRAS.277..758S}.

The viscous torque density, responsible for disc accretion, is given by
\begin{equation}
\frac{\partial T_{\rm vis}}{\partial R}=\frac{\partial}{\partial R}\left(2\pi \nu R^3 \Sigma \frac{\partial \Omega}{\partial R} \right),\label{tvis}
\end{equation} 
where $\nu$ is the viscous shear parameter. The direction of $T_{\rm tid}$ and $T_{\rm vis}$ is opposite, with the balance between the two responsible for the opening of a gap or even a cavity in the disc. 

Even though equations (\ref{ttid}) and (\ref{tvis}) were developed for $q\ll1$, this approximation can provide insights on scaling laws also for higher mass ratios ($q\sim 1$).
Assuming an $\alpha$-prescription by \citet{1973A&A....24..337S} for the viscous shear parameter $\nu=\alpha c_{\rm s} H$, where $\alpha$ is a dimensionless scale parameter, $c_{\rm s}$ is the sound speed and $H$ the disc vertical displacement, we notice that both these torque terms scale with the disc aspect-ratio $H/R$: in particular, integrating equations (\ref{ttid}) and (\ref{tvis}) over the disc, one obtains $T_{\rm tid}\propto (H/R)^{-3}$ and $T_{\rm vis}\propto (H/R)^2$.

Although 1D models predict that no material can cross the gap/cavity edge \citep{2010PhRvD..82l3011L}, 2D and 3D numerical simulations showed that material is able to stream inside the cavity with accretion rates comparable or even higher to those predicted for single objects even in presence of equal mass binaries, that should provide the most intense tidal torques.

However, even though it is widely believed that the accretion rate in binary systems is not affected much by the presence of tidal torques, a suppression similar to that predicted by 1D models may occur when the viscous torque weakens for example when the disc becomes thinner.
The disc internal angular momentum transport mechanisms are more effective for hot/thick discs than for thin/cold ones, both in simple $\alpha$-models \citep{1973A&A....24..337S} and in physically based mechanisms, such as MRI and gravitational instability. 

It is important to mention that the formation of streams of material have been credited to pressure effects \citep{1997ASPC..121..505L}, since pressure can be thought as an alteration to the effective gravitational potential, allowing the mass to overcome the tidal barrier as a consequence of the conservation of the Bernoulli constant along the streamlines. This pressure effect has been recently confirmed by \citet{2016MNRAS.tmp..577D}, even though it becomes relevant only for $H/R\gtrsim 0.1$ in equal mass ratio binary systems.

\section{Numerical Simulations}\label{numsim}

We performed a set of 3D SPH (Smoothed Particle Hydrodynamics) simulations using \textsc{phantom} \citep{2010MNRAS.405.1212L,2010MNRAS.406.1659P,2012JCoPh.231..759P}, varying the disc thickness. We simulated both the binary case and the single central object case, in order to provide a consistent reference for the accretion rate.

\subsection{Initial conditions}\label{initsetup}

Our initial conditions consist of two binary sink particles (able to accrete gas particles), and a finite circumbinary disc of $N_{\rm part}=2\times 10^6$ gas particles in most cases. 
Beside the fluid dynamical forces produced by viscosity and pressure, for which we refer to \citet{2010MNRAS.405.1212L}, the gas particles feel the acceleration produced by the sinks (e.g. \citealp{2013MNRAS.434.1946N}). Each sink exerts on the $i$-th particle the acceleration $\bm f_{i,{\rm pot}}$
\begin{equation}
\bm f_{i,{\rm pot}}=\frac{GM_n(\bm r_n-\bm r_i)}{|\bm r_n-\bm r_i|^3},
\end{equation}   
where $n=1,2$ indicates quantities related to the primary or the secondary object respectively, $G$ is the universal constant of gravitation while $\bm r_1$ and $\bm r_2$ are the positions of the two sink particles of mass $M_1$ and $M_2$ respectively. No smoothing to the potential has been applied as particles are considered accreted when the condition $|r_{1,2}-r_a|<r_{\rm sink}$ is satisfied, where the parameter $r_{\rm sink}=0.05$ is the sink radius, and their kinetic energy is not sufficient to escape the potential well \citep{1995MNRAS.277..362B}. Note that, contrary to previous investigations, we do not prescribe the sink particles on fixed orbits: their motion is determined by the gravitational potential that one exerts on the other and by the back-reaction they receive from the interaction with the gas particles. We neglect the gas-gas gravitational interaction, i.e. no disc self-gravity. 

We choose an equal mass ($q=1$), circular (eccentricity $e=0$) binary system in order to simplify the comparison with the literature. We use code units such that the binary orbital frequency $\Omega_{\rm bin}=1$ (see appendix \ref{codeunitsec}). The masses of the black holes in code units are thus $M_1=M_2=0.5$, the binary separation is $a=1$ and we set the initial velocities of the sinks to obtain circular Keplerian orbits. 

The gas disc is set up by placing particles between an inner radius $R_{\rm in}=2.6 a$ and an outer radius $R_{\rm out}=5a$, where $a$ is the binary separation, in order to obtain an initial surface density distribution of the type $\Sigma=\Sigma_0 R^{-p}$, where $p=2$ and $\Sigma_0$ is chosen in order to have a disc mass $M_{\rm disc}=0.005$, using an initial Monte Carlo particle placement. The initial vertical position of each particle is chosen from a gaussian distribution with standard deviation $H=c_{\rm s}/\Omega$, where $c_{\rm s}$ is the sound speed (see the next Section) and $\Omega$ the Keplerian frequency for a central mass $M_{\rm tot}$.
The velocity of each particle is Keplerian corrected to account for pressure, given by
\begin{equation}
v^2_i=\frac{GM_{\rm tot}}{R_i}-c_{{\rm s},i}^2\left(p+\frac{3}{2}+\ell\right),\label{velinit}
\end{equation}
where $R_i$ is the distance of the $i$-th particle from the centre of mass of the system, $c_{{\rm s},i}$ is the sound speed for the $i$-th particle and $\ell=1/2$ is the power law index for the sound speed.

\subsection{Equation of state and temperature profile}

We to prescribe a locally isothermal equation of state, in order to keep the disc temperature constant in time through the entire length of the simulation
\begin{equation}
P=\frac{k_BT}{\mu m_p}\rho=c_{\rm s}^2\rho,\\
\end{equation}
where $c_{\rm s}^2$ is the squared sound speed of the gas.
The temperature of the gas is then prescribed through the gas sound speed, for which we use the formulation \citep{2014ApJ...783..134F}
\begin{equation}
c_{\rm s}=\frac{H}{R}\left(\frac{GM_1}{R_1}+\frac{GM_2}{R_2}\right)^{\ell}, \label{eos14}
\end{equation}
that implies a constant $H/R$ throughout the disc. This prescription has the nice property of becoming a radial power law around each sink, since equation (\ref{eos14}) reduces to
\begin{equation}
c_{\rm s}=\begin{cases}
\frac{H}{R}v_{1,\rm K}, &\quad {\rm for} \; R_1\ll R_2,\\
\frac{H}{R}v_{2,\rm K}, &\quad {\rm for} \; R_2\ll R_1,\\
\frac{H}{R}v_{\rm K}, &\quad {\rm for} \; R_1 \sim R_2\gg a,
\end{cases}
\end{equation}
\noindent where $a$ is binary separation while $v_{1,\rm K}$, $v_{2,\rm K}$, $v_{\rm K}$ are the Keplerian velocity around the primary, the secondary and and the binary objects respectively.

To compare different disc temperatures we perform simulations using seven different values of $H/R=\{0.13;0.12;0.1;0.08;0.06;0.04;0.02\}$. As mentioned above, the case $H/R=0.1$ is the most used in recent literature \citep{2008ApJ...672...83M, 2013MNRAS.436.2997D, 2014ApJ...783..134F, 2015ApJ...807..131S} and will be useful for a comparison with previous results.

\subsection{Viscosity}\label{viscosec}

SPH employs an artificial viscosity term, in order to resolve shocks. This term acts as a source of viscous diffusion and therefore can be used to model the angular momentum transport in the disc \citep{2010MNRAS.405.1212L}. However, some preliminary tests using artificial viscosity to model the disc viscosity showed an unwanted increase in the accretion rate on to the sinks. This occurs because the artificial shear viscosity $\nu_{\rm AV}$ is dependent on the density $\rho$ of the fluid such that $\nu_{\rm AV}\propto h/H\propto \rho^{-1/3}$, where $h$ is the SPH smoothing length. This implies that if the disc is characterized by strong density gradients, the viscous effects are subject to strong changes throughout the disc, affecting the reliability of the accretion rate.
We therefore decide to introduce just the bare minimum amount of artificial viscosity, using the \citet{1997JCoPh.136...41M} switch with $\alpha_{\rm AV,min}=0.1$, $\alpha_{\rm AV,max}=0.5$ and $\beta_{\rm AV}=2$ everywhere to prevent particle interpenetration.

 To compute the ``physical'' viscosity we use the implementation of Navier-Stokes viscosity similar to that given by \citet{1994ApJ...431..754F}, as described in \citet{2010MNRAS.405.1212L} (see their section 3.2.4).
We set the bulk viscosity $\zeta=0$, while $\nu$ is computed using an $\alpha$-prescription that reduces to the standard \citet{1973A&A....24..337S} prescription around each sink:
\begin{align}
\nu&=\alpha_{\rm SS} c_{\rm s} \frac{H}{R}f, \label{viscpresc}\\
f&=\min\left(R_1,R_2\right),
\end{align}
where $c_{\rm s}$ is given by equation (\ref{eos14}), $R_1$ and $R_2$ are the distances from $M_1$ and $M_2$ respectively; we set $\alpha_{\rm SS}=0.1$. Our choice of $\alpha_{\rm SS}$ is equal to \citet{2014ApJ...783..134F} and is consistent with the equivalent value extrapolated from MHD simulations by \citet{2012ApJ...749..118S,2015ApJ...807..131S}, although this result needs to be validated for thinner discs. Note that the main angular momentum transport process at $\approx 0.1\, {\rm pc} $ separation in an AGN disc might be associated with gravitational instabilities \citep{2003MNRAS.339..937G,2012AdAst2012E..11L}. For the thin discs in AGN, self-gravitating angular momentum transport is local and is expected to provide equivalent $\alpha \approx 0.1-0.3$, consistent with our choice \citep{2009MNRAS.393.1423C}.
 
Parameters employed for all the simulations in this study are listed in Table \ref{tabsim}.

\subsection{Resolution}\label{resosec}
Beside the simulations discussed in the next Sections (those denoted by ${\rm S}$ in Table \ref{tabsim}), we performed a set of simulations (denoted by ${\rm Reso}$ in Table \ref{tabsim}) varying $N_{\rm part}$ as a convergence test. The results for $N_{\rm part}=1.0\times 10^6$ converge to the higher resolution simulations. The scale height is well resolved throughout the disc for every $H/R$ (see Table \ref{tabsim}). However, the cavity region is obviously poorly resolved due to its low density. While in most cases the disc thickness is resolved even in the cavity, for $H/R=\{0.04;0.02\}$ $h/H$ can be much above unity ($h/H\sim \{6;8\}$, respectively). This results in an increase of the artificial viscosity inside the cavity for $H/R=\{0.04;0.02\}$. 

We estimate this excess in the cavity region by computing the value of $\alpha_{\rm SS,AV}$ corresponding to the artificial viscosity using \citep{2010MNRAS.405.1212L}
\begin{equation}
\alpha_{\rm SS,AV}=\frac{1}{10}\alpha_{\rm AV,max}\frac{h}{H},
\end{equation}
through which we are able compare the magnitude of the artificial viscosity with respect to the ``physical'' one.
The values of $\alpha_{\rm phys,AV}$ in the cavity region for the cases $H/R=\{0.04;0.02\}$ are $\alpha_{\rm SS,AV}\sim 0.3$ and  $\alpha_{\rm SS,AV}\sim 0.4$, respectively, to be compared to the value of $\alpha_{\rm SS}=0.1$ that we prescribed for the physical viscosity. Thus, for the two thinnest cases the evolution of the gas within the cavity is dominated by numerical effects.

\begin{table*}
\centering
\begin{tabular}{|l|cccccccc|}
\hline
&$H/R$&$N_{\rm part}$&$\alpha$&$\nu$&$t_\nu/t_{\rm bin}$&$\langle h/H\rangle $&SO\\
\hline
S1&$0.13$&$2.0\times 10^6$&$0.10$&$1.7\times 10^{-3}$&$701$&$0.09$&yes\\
S2&$0.12$&$2.0\times 10^6$&$0.10$&$1.4\times 10^{-3}$&$823$&$0.09$&yes\\
S3&$0.10$&$2.0\times 10^6$&$0.10$&$1.0\times 10^{-3}$&$1185$&$0.11$&yes\\
S4&$0.08$&$2.0\times 10^6$&$0.157$&$1.0\times 10^{-3}$&$1185$&$0.13$&no\\
S5&$0.08$&$2.0\times 10^6$&$0.10$&$6.4\times 10^{-4}$&$1851$&$0.13$&yes\\
S6&$0.06$&$2.0\times 10^6$&$0.177$&$6.4\times 10^{-4}$&$1851$&$0.15$&no\\
S7&$0.06$&$2.0\times 10^6$&$0.10$&$3.6\times 10^{-4}$&$3291$&$0.15$&yes\\
S8&$0.04$&$2.0\times 10^6$&$0.10$&$1.6\times 10^{-4}$&$7406$&$0.20$&yes\\
S9&$0.02$&$2.0\times 10^6$&$0.10$&$4.0\times 10^{-5}$&$29625$&$0.32$&yes\\
\hline
Reso1&$0.10$&$1.5\times 10^6$&$0.10$&$1.0\times 10^{-3}$&$1185$&$0.12$&no\\
Reso2&$0.10$&$1.0\times 10^6$&$0.10$&$1.0\times 10^{-3}$&$1185$&$0.14$&no\\
Reso3&$0.10$&$5.0\times 10^5$&$0.10$&$1.0\times 10^{-3}$&$1185$&$0.17$&no\\
Reso4&$0.08$&$1.0\times 10^6$&$0.10$&$6.4\times 10^{-4}$&$1851$&$0.16$&no\\
Reso5&$0.06$&$1.0\times 10^6$&$0.10$&$3.6\times 10^{-4}$&$3291$&$0.19$&no\\
Reso6&$0.04$&$1.0\times 10^6$&$0.10$&$1.6\times 10^{-4}$&$7406$&$0.25$&no\\
Reso7&$0.02$&$1.0\times 10^6$&$0.10$&$4.0\times 10^{-5}$&$29625$&$0.40$&no\\
Reso8&$0.02$&$5.0\times 10^5$&$0.10$&$4.0\times 10^{-5}$&$29625$&$0.51$&no\\
\hline
\end{tabular}
\caption{Summary of the simulations. For each simulation we give the aspect ratio $H/R$, the number of particles $N_{\rm part}$, the $\alpha$ viscous parameter, the shear parameter $\nu$, the viscous time $t_{\nu}/t_{\rm bin}$ (equation \ref{tnusim}) expressed in binary orbital periods units and the initial smoothing-length normalized to the disc thickness (disc average). The column SO is to indicate if the corresponding single object simulation has been performed. Notice that the value reported for $\nu$ is computed at $R_1=R_2=a$. For completeness, the parameters in each simulations are: inner disc radius $R_{\rm in}=2.6$, outer disc radius $R_{\rm out}=5$, total binary mass $M_{\rm tot}=1$, disc mass $M_{\rm disc}=0.005$, density power law index $p=2$, sound speed power law index $\ell=0.5$, binary mass ratio $q=1$, and binary orbital eccentricity $e=0$.}\label{tabsim}
\end{table*}

\section{Results}\label{resu} 

The finite size of our discs causes spreading to larger radii as time passes, causing the accretion rate to vary. The evolution time-scale for accretion discs is given by the viscous time $t_\nu$
\begin{align}
t_\nu&=\frac{2}{3}\frac{R^2_{\rm out}}{\nu(R_{\rm out})}=\frac{5^{3/2}}{3\pi\alpha} \left(\frac{H}{R}\right)^{-2}t_{\rm bin}\\
&\approx 11.85\cdot \left(\frac{H}{R}\right)^{-2}t_{\rm bin} \label{tnusim},
\end{align}
where $t_{\rm bin}$ is the binary orbital period
\begin{equation}
t_{\rm bin}=\frac{2\pi}{\Omega_{\rm bin}}=2\pi\left[\frac{G(M_1+M_2)}{a^3}\right]^{-1/2}.\label{tbin}
\end{equation}
Any comparison between different regimes of disc thickness needs therefore to be done at the same $t/t_{\nu}$.
The time for the comparison has to be chosen to be long enough to allow the achievement of quasi-stationarity, but much smaller than $t_\nu$ in order to prevent excessive relaxation of the initial conditions. Indeed, as will be seen below, circumbinary disc properties and structure evolve differently as the time passes and a comparison at later times would not be reliable anymore.

We compare our simulations at times which are sufficiently long to overcome the initial transient and computationally tractable for the cases we simulated. However, where possible we let the discs evolve up $t\sim 0.2\, t_\nu$. 

Due to its long viscous time, the $H/R=0.02$ calculation reached just $t=0.07 \, t_\nu\sim 2000 \, t_{\rm bin}$; however we believe that the results for this case provide a valid estimate for the accretion rate and for this reason should not be discarded.  

\subsection{Single object simulations}\label{singcensec}

In addition to the simulations of binary systems, we have also performed seven reference simulations around a single object (marked by ${\rm SO=yes}$ in Table \ref{tabsim}), one for each value of $H/R$ used in the binary simulations.

We use the reference simulations to evaluate the accretion on the single central object $\dot M_0$ and thus renormalize the results obtained for the binary case.
The initial setup of the disc of each reference run is the same as that described for the binary case in Section \ref{initsetup}. The only differences are the presence of a Keplerian potential produced by a mass $M=M_1+M_2$ in the centre of mass of the system and a sink radius $R_{\rm sink,SO}=1$. The number of particle used was $N=2\times 10^6$.

\begin{figure}
\includegraphics[width=0.47\textwidth]{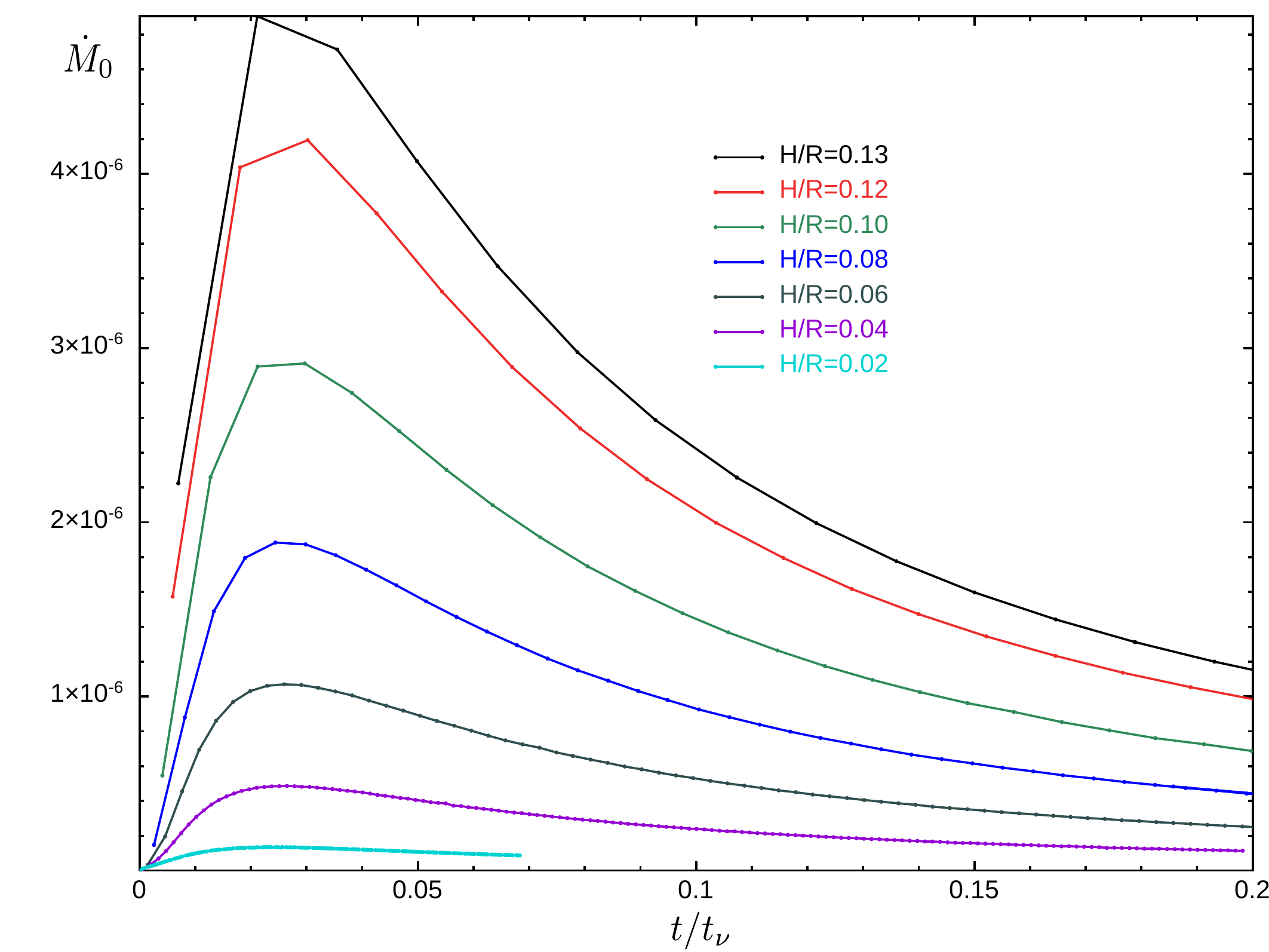}
  \includegraphics[width=0.47\textwidth]{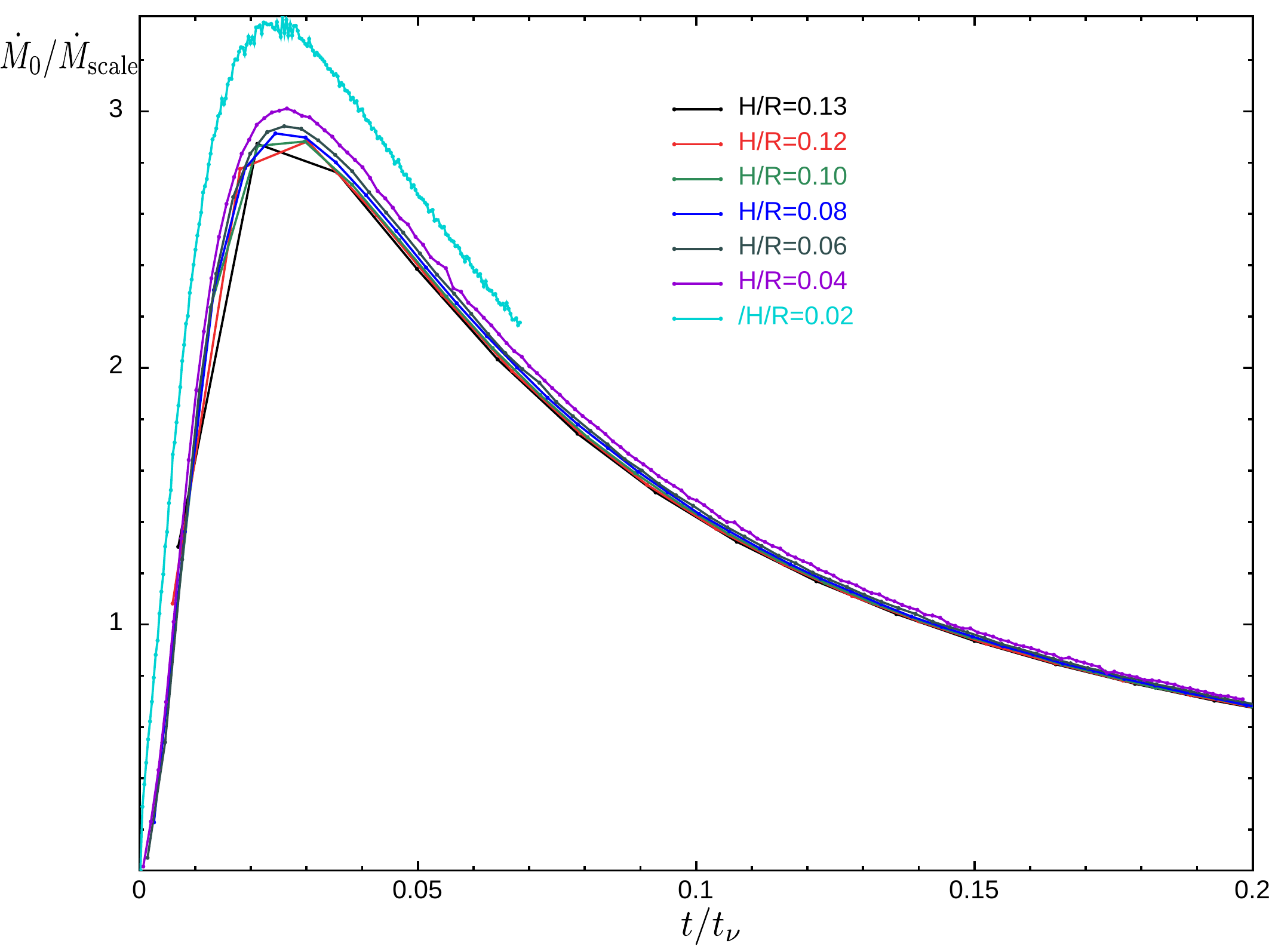}
\caption{Top panel: absolute accretion rate $\dot M_0$ around a single object in code units, averaged over $10\, t_{\rm dyn}$, at the sink radius $R=R_{\rm sink}=1$, as a function of $t/t_\nu$. The different colours represents the seven cases of $H/R=\{0.13;0.12;0.10;0.08;0.06;0.04;0.02\}$ (black, red, green, blue, grey, purple, cyan lines, respectively) we studied.
Bottom Panel: $\dot M_{\rm bin}/\dot M_{\rm scale}$, where $\dot M_{\rm scale}$ normalization constant given by equation (\ref{normc}), again as a function of $t/t_\nu$. It should be noticed that the absolute value of the accretion rate decreases, for lowering thicknesses, as $(H/R)^2$. In particular, from the top panel it can be easily noticed a difference of a factor $\sim 25$, going from $H/R=0.1\rightarrow0.02$.}
\label{isodisc}
\end{figure}

In the upper panel of Figure \ref{isodisc} the results for the accretion rate on to the single central object $\dot M_0$ (in code units) as a function of $t/t_\nu$ are plotted: the evolution of the time-varying accretion rate occurs on a viscous time $t_\nu$. In the lower panel of Figure \ref{isodisc}, the accretion rate is renormalized to $\dot M_{\rm scale}$ given by
\begin{equation}
\dot M_{\rm scale}=3\pi \nu(R_{\rm out})\Sigma(R_{\rm out})\propto (H/R)^2,\label{normc}
\end{equation}
which gives an analytical estimate of the order of magnitude of the accretion rate. The accretion rate is obtained computing how many particles are accreted at $R_{\rm sink,SO}$. 

The accretion rates plotted in Figure \ref{isodisc} are characterized by an initial transient at the time $t\approx0.025\,t_\nu$ for each $H/R$. After that, the disc reaches quasi-stationarity and the accretion rate lowers as the time passes due to the spread of the disc toward larger radii.

The lower panel of Figure \ref{isodisc} shows that the accretion rates from our simulations scale as predicted by theory. The only exception is our thinnest case with $H/R=0.02$ that overestimates the accretion rate with respect to $\dot M_{\rm scale}$. Due to the higher value of $\langle h/H\rangle$ throughout the disc, the artificial viscosity gives a non-negligible contribution with respect to the physical one resulting in a slight overestimate of the accretion rate in this case. This effect is less visible but already present in the other regimes. It should be noticed that the lowest curve is the thickest case $H/R=0.13$ (that shows the lowest $\langle h/H\rangle$), while $\dot M_0/\dot M_{\rm scale}$ progressively grows as $\langle h/H\rangle $ increases for thinner regimes. 

The lower panel in Figure \ref{isodisc} confirms that the outcome of the simulations for $\dot M_0$ are consistent with the analytical predictions both for the scaling, i.e. $\dot M_0\propto (H/R)^2$, and for the magnitude
\begin{equation}
\dot M_{0,{\rm theor}}\approx 3\pi \nu \Sigma.\label{dotmteor}
\end{equation}

\subsection{Binary simulations}\label{binsimsec}

\begin{figure*}
\centering
\includegraphics[width=\textwidth]{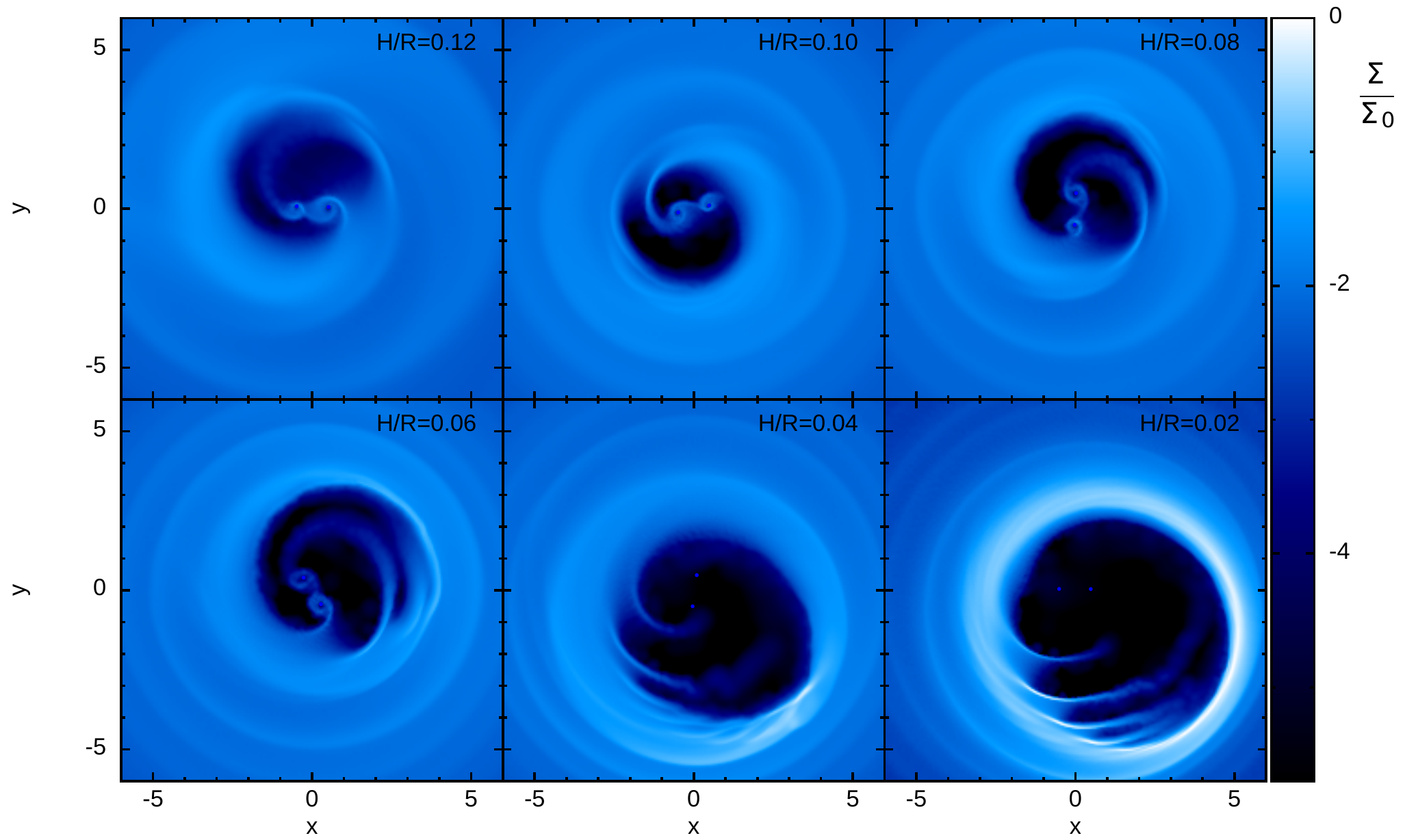}
\caption{Colour plots of the column density normalized to $\Sigma_0$ of our simulations (logarithmic scale). From left on the top row $H/R={0.12;0.1;0.08}$, bottom row $H/R={0.06;0.04;0.02}$. Snapshots were taken at times $t\sim 0.2 t_\nu$ in any case, except $H/R=0.02$ which was taken at time $t\sim0.07 t_\nu$.}
\label{colourdensplot}
\end{figure*}

Figure \ref{colourdensplot} shows column density in our binary systems after $t\sim0.2t_\nu$ for $H/R=\{0.12;0.1;0.08;0.06;0.04\}$ and at time $t\sim0.07t_\nu$ for the case $H/R=0.02$. We observe the formation of circumprimary and circumsecondary discs around each sink for the cases $H/R=\{0.12;0.1;0.08;0.06\}$ but not for the cases $H/R=\{0.04;0.02\}$. 
This is probably a numerical effect related to the previously discussed increase of the ratio $h/H$ inside the cavity, which results in an unwanted excess of the artificial viscosity in the two thinnest cases.

The size of the cavity is larger for low $H/R$. This is expected. The criteria for gap opening state that, for lower $H/R$, the gap can be opened by progressively less massive binary companions \citep{1993prpl.conf..749L,2006Icar..181..587C}. This occurs because the viscous torques, that are responsible for driving the gas inward, scale as $T_{\rm visc}\propto\nu\propto (H/R)^2$, while the tidal torques, responsible for the ``dam'' effect at the edge of the cavity, scale as $T_{\rm tid}\propto (H/R)^{-3}$. The radius at which the two contrasting torque contributions are equal  approximately gives the truncation radius of the disc (for a more accurate discussion see \citealp{1994ApJ...421..651A}).

Figure \ref{colourdensplot} also shows that each simulation has developed an eccentric cavity, characterized by an overdense lump of material at its edge. This effect has been widely observed in the literature \citep{2001A&A...366..263P, 2006ApJ...652.1698D, 2006A&A...447..369K, 2008ApJ...672...83M, 2013MNRAS.436.2997D, 2014ApJ...783..134F, 2015MNRAS.448.3545D}, interpreted as being due to the unstable growth of spontaneous deviations of the gas from circular motion, starting from super-hump theory \citep{1991ApJ...381..259L}. This interpretation relates to the size of the cavity and, more specifically, to whether the resonances believed to damp this instability fall in the disc region or not, as suggested by \citet{2001A&A...366..263P}. However, we caution that even though this interpretation applies to low mass ratios ($10^{-3}\leq q\leq 3\cdot 10^{-2}$ \citealp{2001A&A...366..263P, 2006ApJ...652.1698D, 2006A&A...447..369K}), the extension of the model to higher companion masses is not straightforward and still needs further investigation. 
\citet{2016MNRAS.tmp..577D} pointed out that a transition between circular and lopsided discs occurs for mass ratios $q>0.04$ as a consequence of the loss of stable orbits across the corotation region in the restricted three body problem (orbits around L4 and L5 Lagrangian points) in that range of masses, and other viscous effects. 

It can be also noticed that the prominence of the lump appears to be influenced by the disc thickness, in particular the overdensity is more marked when the disc is thinner.

\subsection{Suppression of accretion for thin discs}\label{tidaldam}

\begin{figure*}
\includegraphics[width=0.47\textwidth]{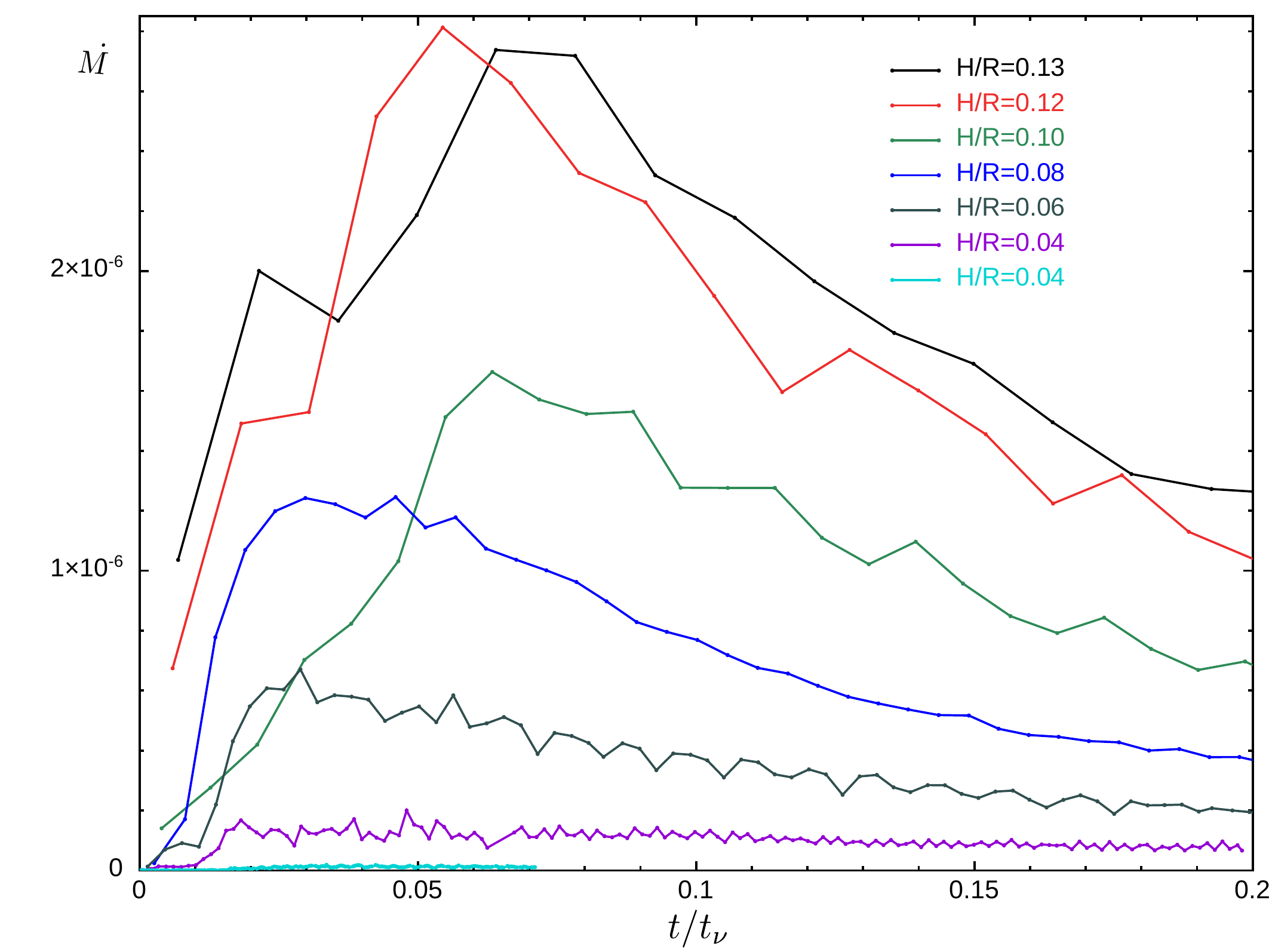}
  \includegraphics[width=0.47\textwidth]{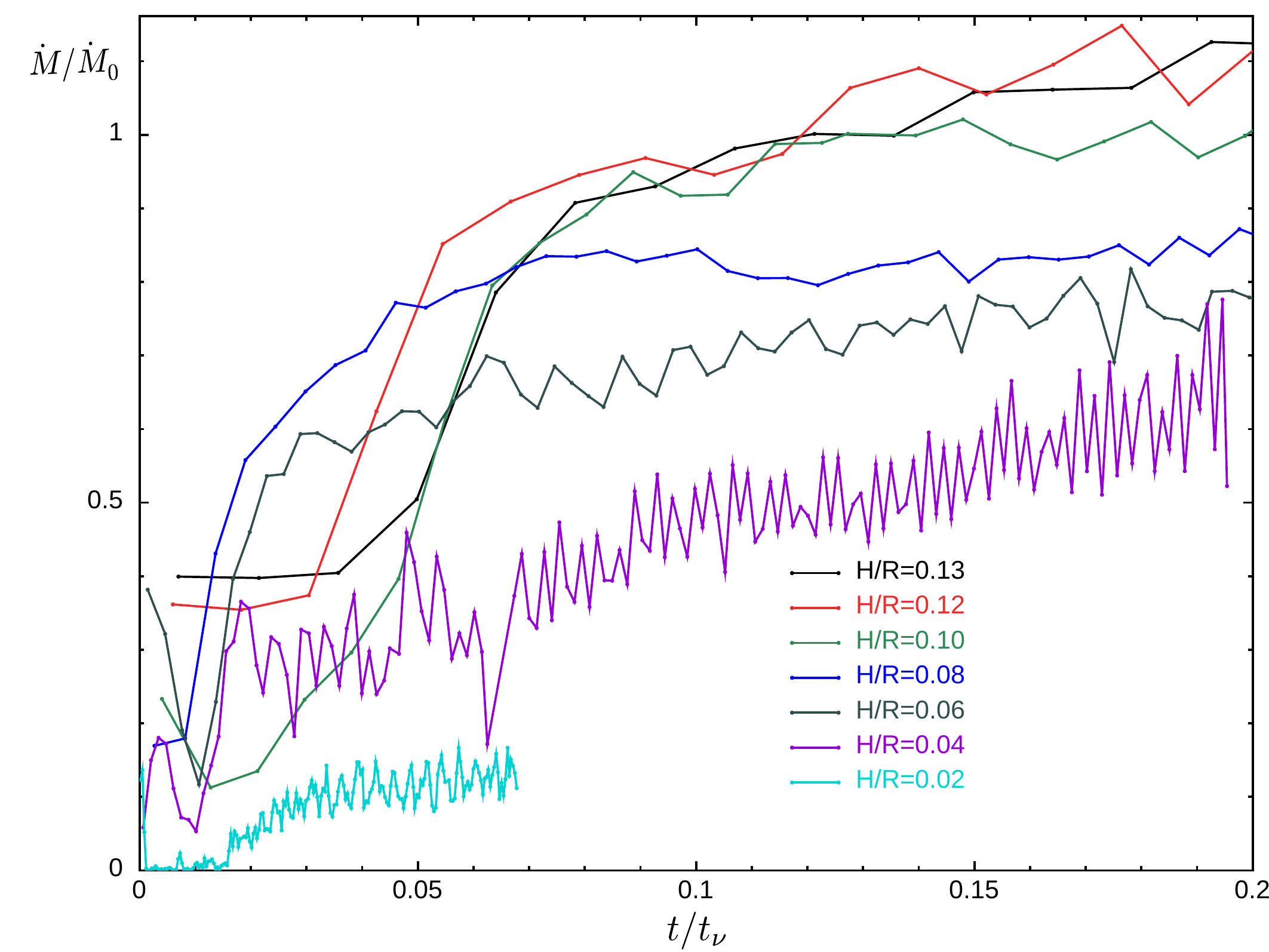}
\caption{Left panel: y-axis accretion rate in code units, each point in the plot obtained averaging $\dot M_{\rm bin}$ over $10 t_{\rm bin}$, as a function of $t/t_{\rm bin}$. Right panel:  $\dot M_{\rm bin}/\dot M_0$ as a function of $t/t_{\rm bin}$. Different colours refers to different $H/R$ regimes, in particular $H/R=\{0.13;0.12;0.1;0.08;0.06;0.04;0.02\}$ black, red, green, blue, grey, purple, cyan lines respectiely.}
\label{MbMi}
\end{figure*}

Figure \ref{MbMi} shows the accretion rate across the cavity edge $\dot M_{\rm bin}$ for our binary simulations compared to the corresponding value obtained in Section \ref{singcensec} for a single black hole. The left panel shows $\dot M_{\rm bin}$ versus $t/t_\nu$ for the various disc thicknesses. As $H/R$ is reduced, $\dot M_{\rm bin}$ drops significantly. A reduction in $\dot M$ with thickness is expected independently of the presence of the binary (upper panel of Figure \ref{isodisc}). However for our $\alpha$-model, we expect that $\dot M\propto (H/R)^2$. The right panel of Figure \ref{MbMi} shows that the suppression of the mass accretion rate is much stronger than this. The right panel of Figure \ref{MbMi} shows the ratio $\dot M_{\rm bin}/\dot M_0$, and we recall that we have demonstrated (lower panel of Figure \ref{isodisc}) that $\dot M_0 \propto (H/R)^2$. This represents the fraction of the unperturbed gas flow that makes it into the cavity. While for $H/R\sim 0.1$ we recover the known result that $\dot M_{\rm bin}\sim \dot M_0$, implying that the binary does not prevent matter from accreting, the situation changes drastically for lower $H/R$. For $H/R=0.02$, for example, only $\sim 15$ per cent of the unperturbed mass makes it into the cavity. We thus conclude that 
\begin{equation}
\dot M_{\rm bin}=\xi(H/R)\dot M_0,\label{xiscale}
\end{equation}
where $\xi$ is a function of $H/R$. For the thicker cases the value of $\xi(H/R)$ appears to saturate at $\sim 1$ for growing $H/R$. This is in contrast with the results obtained by \citet{2014ApJ...783..134F} and \citet{2015ApJ...807..131S} that found values of $\xi$ above unity ($\xi \sim 1.6$ \citealp{2014ApJ...783..134F}, $\xi \sim 1.4$ \citealp{2015ApJ...807..131S}) for a unitary mass-ratio binary system with $H/R=0.1$, but in agreement with \citet{2013MNRAS.436.2997D} that found $\xi=1.015$ for the case of interest. 

While for large $H/R$ the value of $\dot M_{\rm bin}/\dot M_0$ appears to reach a well defined asymptote, for lower $H/R$ it increases with time. This deserves a brief discussion. In our thinnest cases, the binary accretes much less than the corresponding $q=0$ discs, accumulating some material at the edge of the cavity. This translates into a slower time evolution of the accretion rate which remains almost constant after the peak (see purple and cyan lines in the left panel of Figure \ref{MbMi}), while the value of $\dot M_0$ keeps decreasing; as a consequence, the ratio $\dot M_{\rm bin}/\dot M_0$ increases with time. For this reason the value of $\dot M_{\rm bin}/\dot M_0$ is most reliable when the discs have evolved for enough time to overcome the initial transient, but not so much that the viscous evolution of the system modifies the disc structure\footnote{Recall that our discs do not reach a steady-state since they have a finite mass and radius.}.

The left panel of Figure \ref{MbMi} highlights another important feature. The thickest cases ($H/R=\{0.13;0.12;0.10\}$) show a time-shift in the initial peak. This feature is likely related to our initial conditions for the velocity field of the gas, equation (\ref{velinit}). In the presence of a binary potential our choice underestimates the equilibrium velocity of the gas. The thickest regimes, in which viscous forces are stronger, result then in a less steep initial transient, that reaches a lower maximum. However, this shift does not affect the evolution at later times.
 \begin{table}
\centering
\begin{tabular}{|c|ccccccc|}
\hline
 $H/R$&$0.13$&$0.12$&$0.10$&$0.08$&$0.06$&$0.04$&$0.02$\\
\hline
 $\xi(H/R)$&$0.97$&$0.99$&$0.95$&$0.82$&$0.70$&$0.47$&$0.18$\\
\hline
\end{tabular}
\caption{Values of $\xi(H/R)$ taken averaging $\xi(H/R)$ between $0.07\, t_\nu<t<0.15 \,t_\nu$ ($H/R=\{0.13;0.12;0.10;0.08;0.06;0.04\}$). The value of $\xi$ for the case $H/R=0.02$ is obtained instead averaging between  $0.05\,t_\nu<t< 0.07 \,t_\nu$.}\label{xihor}
\end{table}

\begin{figure}
\centering
\includegraphics[width=0.45\textwidth]{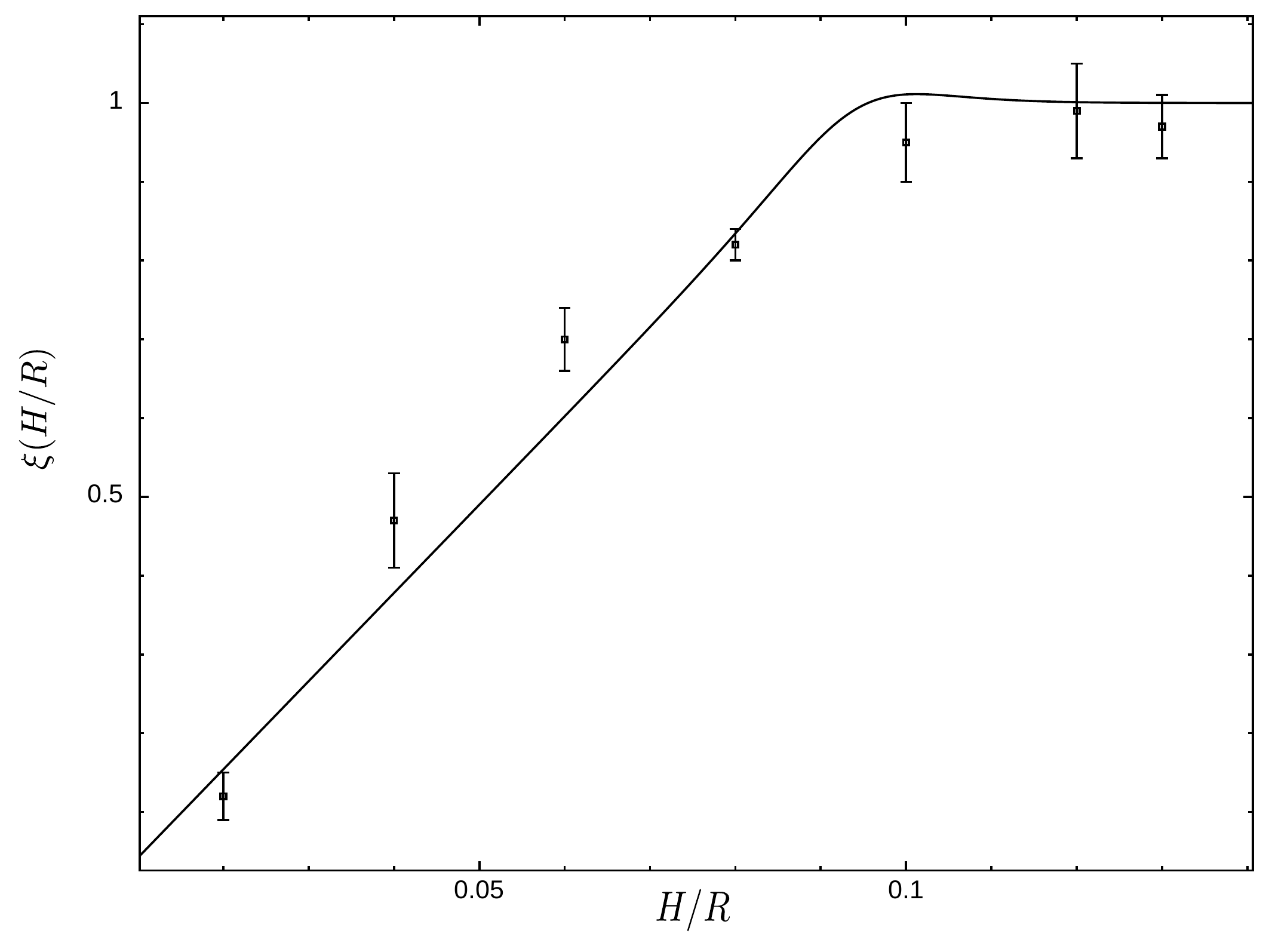}
\caption{Plot of $\xi(H/R)$ as a function of $H/R$, with the values in Table \ref{xihor}. The error bars represent the square root of the mean quadratic error of the values assumed by $\xi(H/R)$ for times between $0.07\, t_\nu<t<0.15 \,t_\nu$ (for the case $H/R=0.02$, $0.05\, t_\nu<t<0.07 \,t_\nu$). The black line plots an interpolating function such as equation (\ref{linpred}).}\label{suppraccr}
\end{figure}

Table \ref{xihor} shows the time average of $\xi$ for each value of the disc thickness, where the average has been taken after the initial transient but before viscous evolution has significantly affected our results. In particular, we generally average between $0.07t_{\nu}$ and $0.15t_{\nu}$, except for $H/R=0.02$, for which we average between $0.05-0.07t_{\nu}$.

Figure \ref{suppraccr} shows the average $\xi$ as a function of $H/R$, where the error bars are the square root of the mean quadratic error in the average procedure. It appears that $\xi$ increases linearly with $H/R$ for $H/R<0.1$, saturating at around unity for $H/R>0.1$. We thus write
\begin{equation}
\xi(H/R)\approx\begin{cases}
10\cdot H/R, &{\rm for} \; H/R\leq 0.1\\
1, &{\rm for} \; H/R> 0.1
\end{cases}
.\label{linpred}
\end{equation}
The black line in Figure \ref{suppraccr} shows an interpolating function with a linear behaviour for $H/R<0.1$ and constant $\xi=1$ for $H/R>0.1$.

The typical values of $H/R$ in AGN discs are believed to span $H/R\sim 10^{-2}-10^{-3}$ \citep{1973A&A....24..337S, 1990A&A...229..292C, 1998ApJ...506L..97N, 2003MNRAS.339..937G, 2012AdAst2012E..11L, 2015MNRAS.451.3941G} depending on the region of the disc examined and on the degree of self-gravity in the disc \citep{2009ApJ...700.1952H}.
If the law in equation (\ref{linpred}) keeps holding also for these physical values of $H/R$, we expect to have $\xi \sim 0.1-0.01$, implying that the accretion rate is suppressed up to a factor $10^2$ with respect to the normal AGN activity.

A similar effect has been recently found during the last phases of supermassive black hole binary mergers \citep{2016MNRAS.457..939C}. In that case tidal torques acts as barriers for the gas during the late gravitational inspiral phase before the black hole merger: the binary companion squeezes the gas of the thin circumprimary disc toward the primary object during its orbital decay, causing an enhancement in the accretion rate and thus predicting a flare in the luminosity of the system just before the binary merger (in contrast, \citealp{2012MNRAS.423L..65B} found that this effect does not occur for thicker discs).

\subsection{Accretion Variability as a function of the disc temperature}\label{resultsectime}

In this Section we analyze the fast variability of the mass flow across the cavity edge, $\dot M_{\rm bin}$, and the accretion rate as computed directly from accretion on to the binary, $\dot M_{\rm sink}$.

\begin{figure*}
\centering
\includegraphics[width=0.40\textwidth]{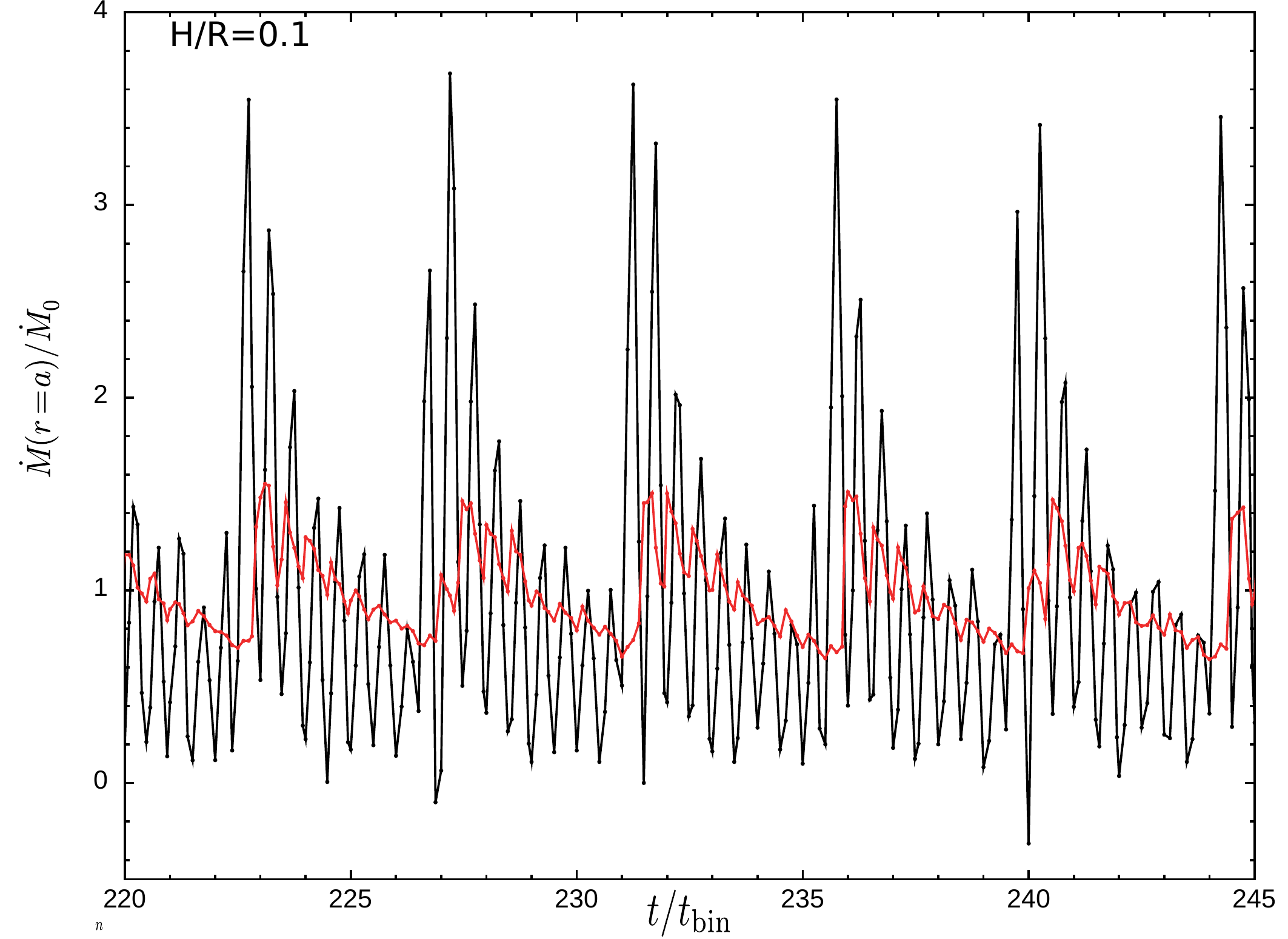}
\includegraphics[width=0.40\textwidth]{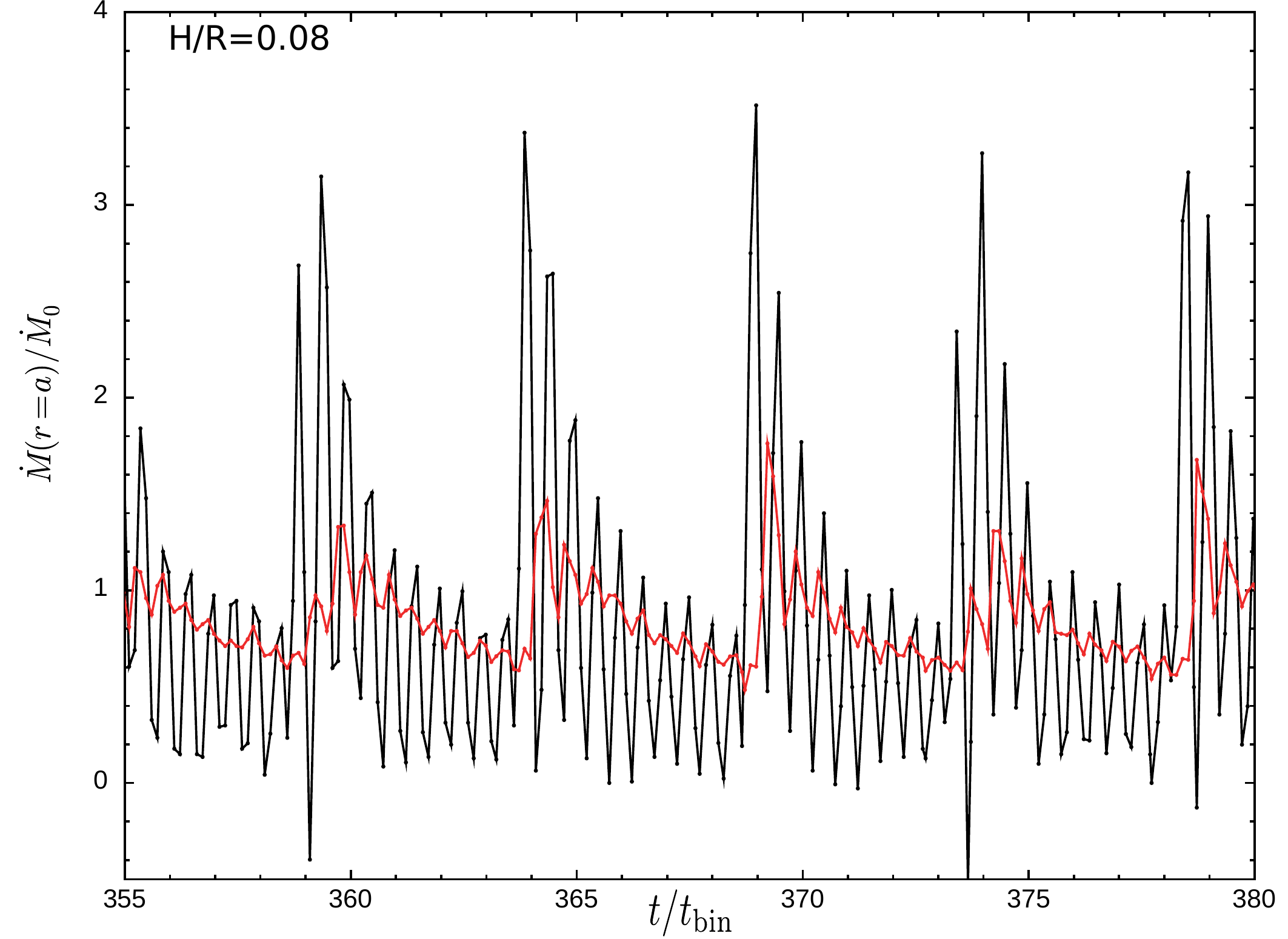}\\
\centering
\includegraphics[width=0.40\textwidth]{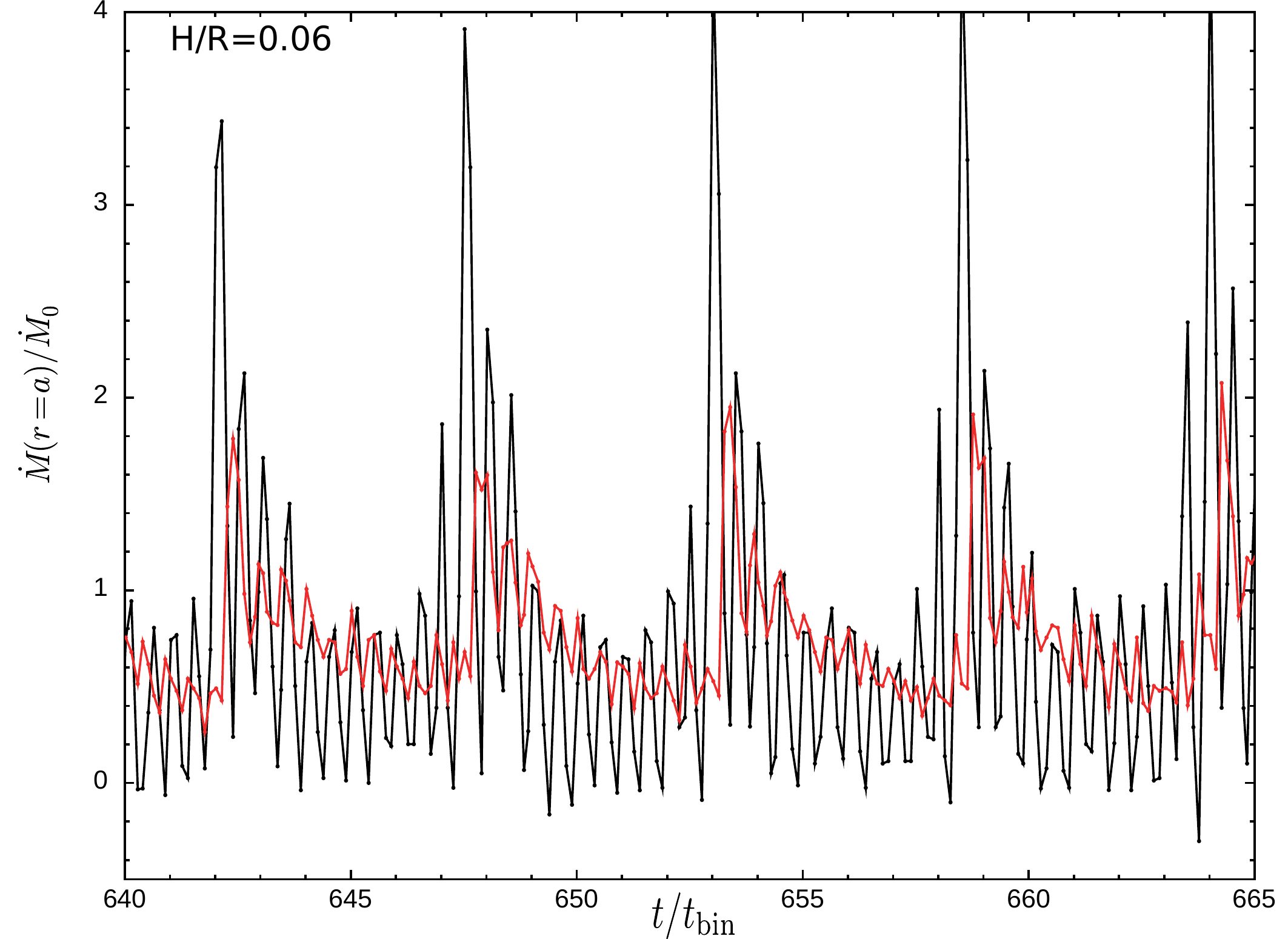}
\includegraphics[width=0.40\textwidth]{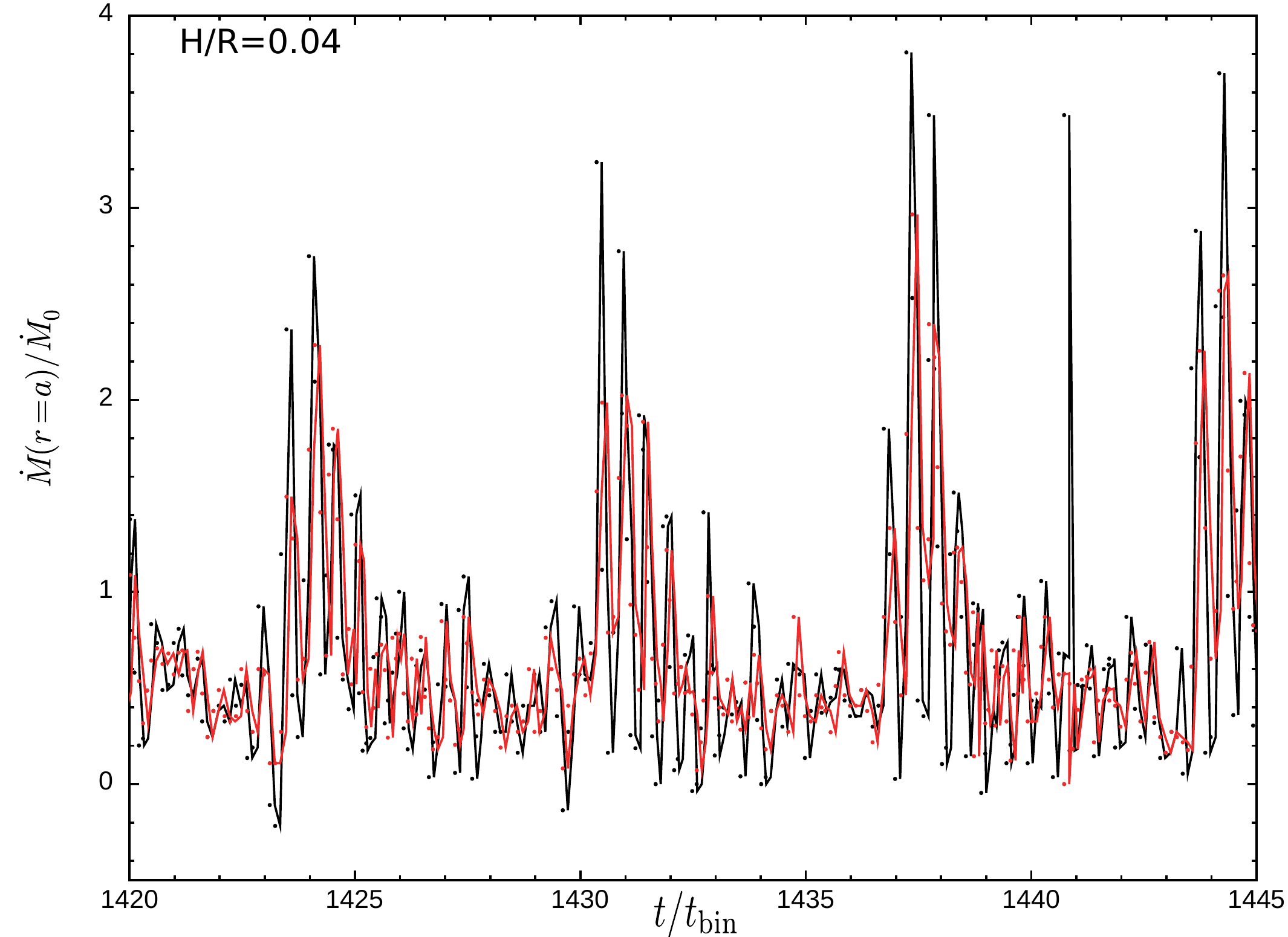}
\caption{From top-left to bottom right: $H/R=\{0.1;0.08;0.06;0.04\}$ respectively. Black line: Time variability of the mass flux across the radius $R=a$, $\dot M_{\rm bin}/M_0$ (normalized to the corresponding averaged $\dot M_0$), and red line: accretion rate on to the binary $\dot M_{\rm bin}/\dot M_0$. The panels show the various regimes at $t\sim 0.2 t_\nu$.}
\label{timevarbin}
\includegraphics[width=0.40\textwidth]{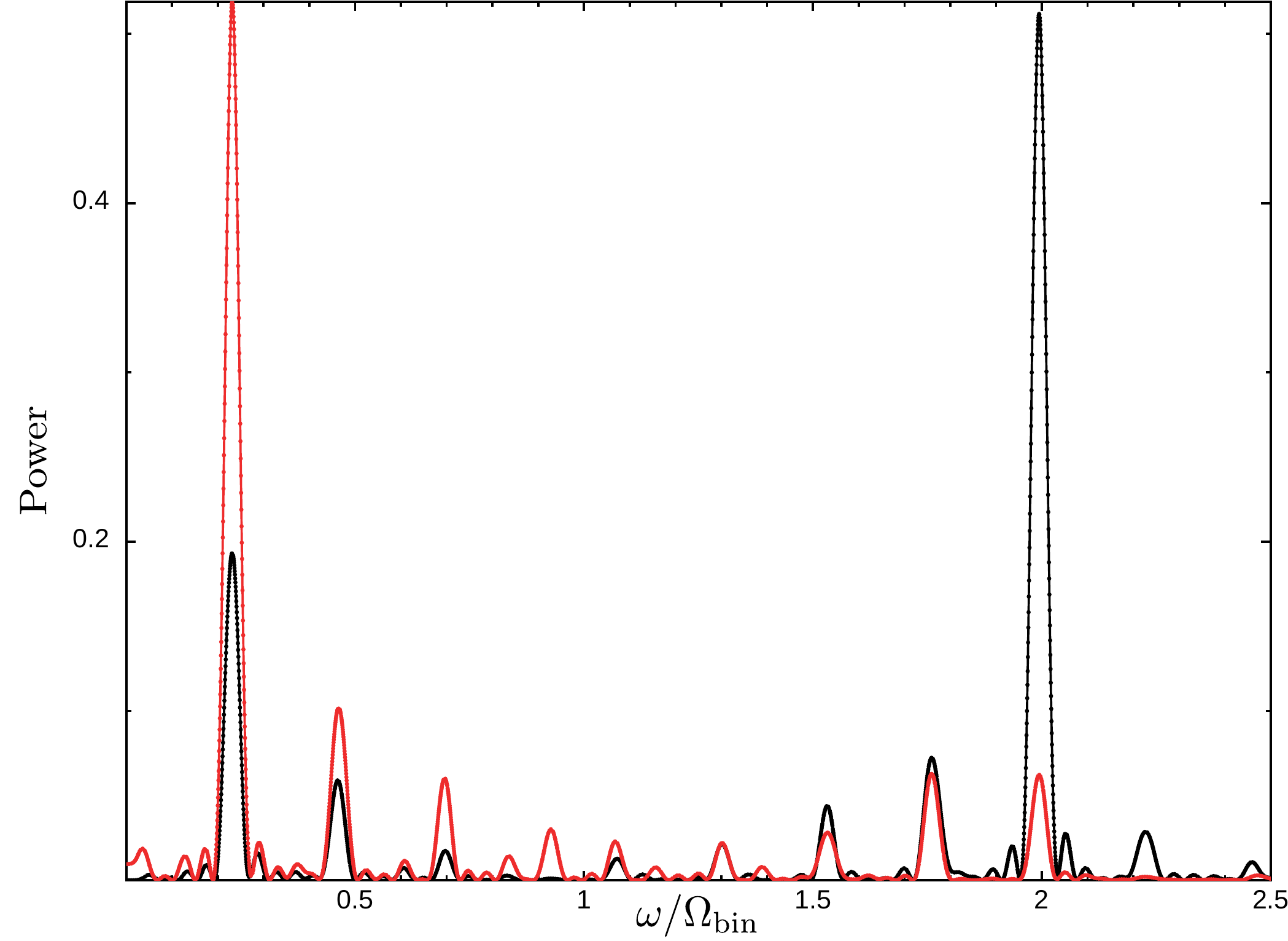}
\includegraphics[width=0.40\textwidth]{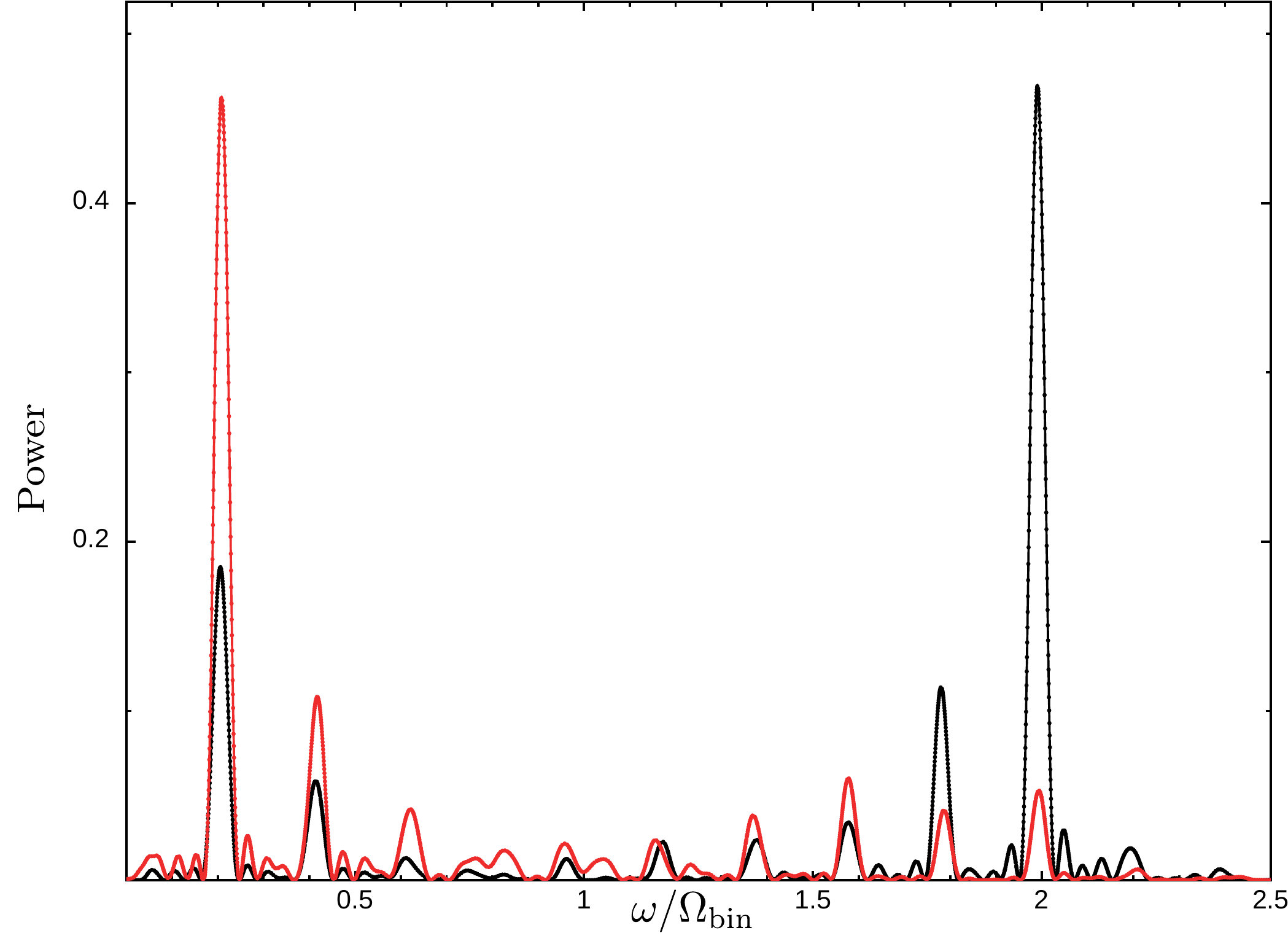}\\ %
\includegraphics[width=0.40\textwidth]{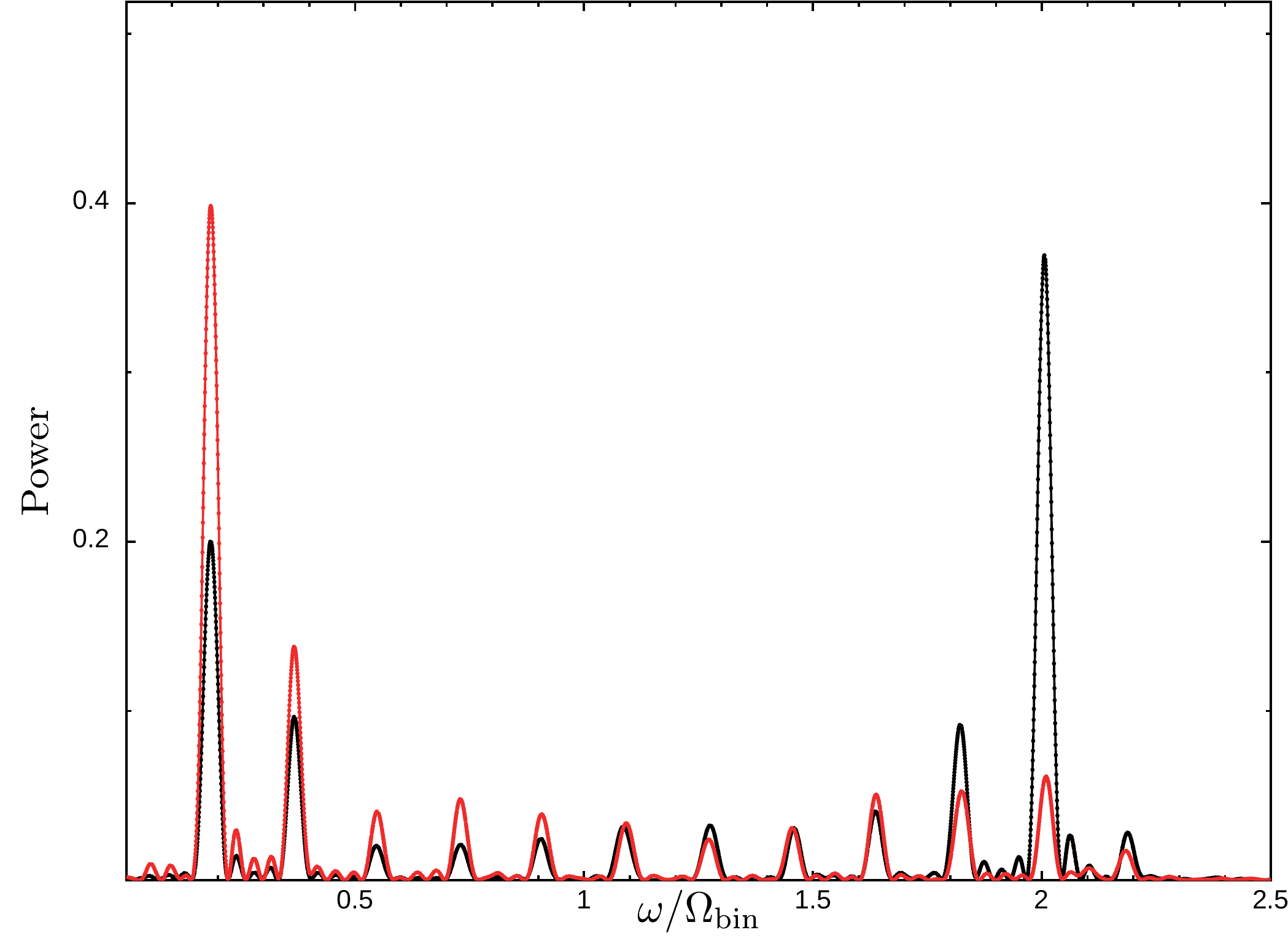}
\includegraphics[width=0.40\textwidth]{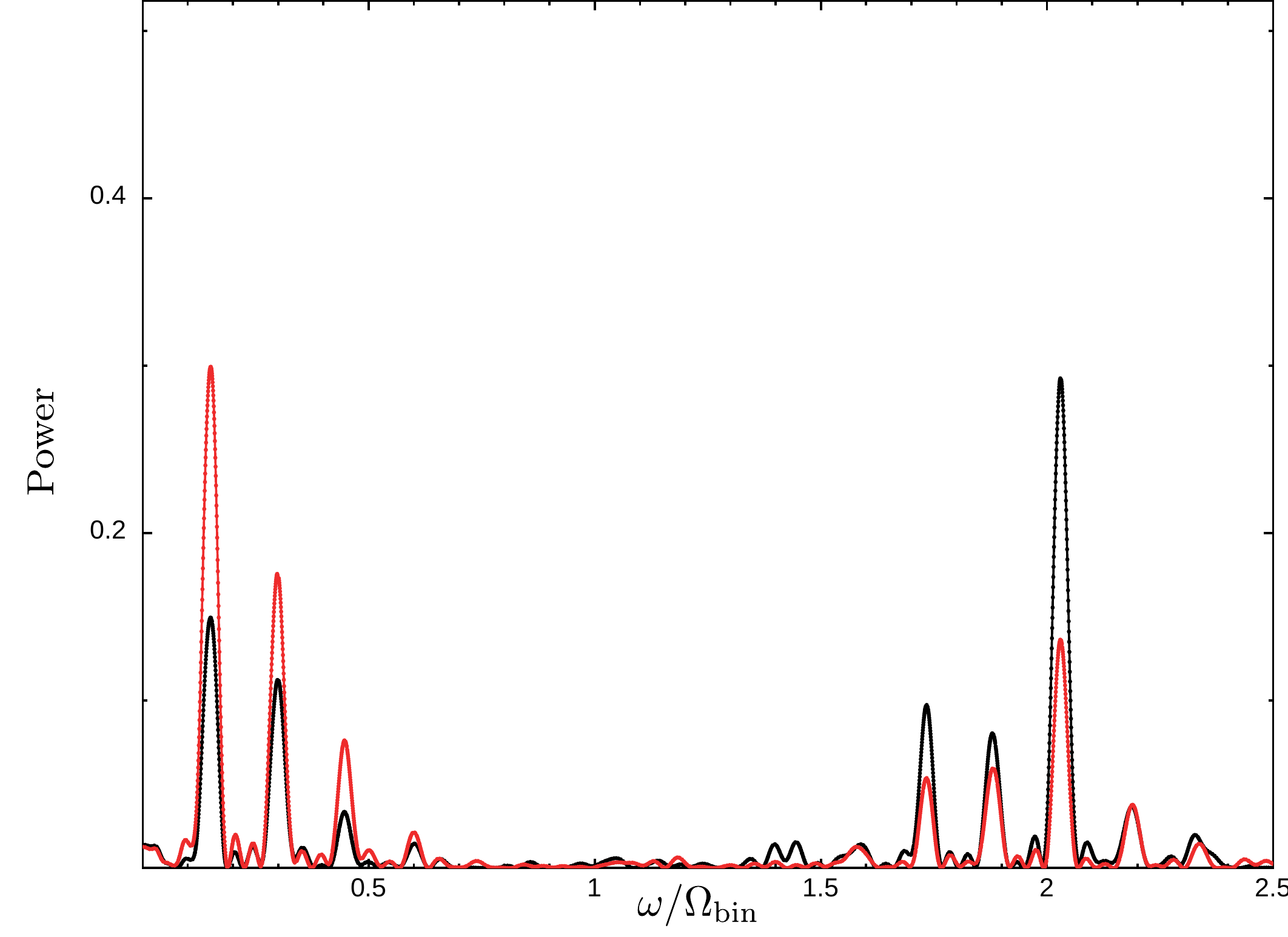}
\caption{From top-left to bottom right: periodograms of the accretion rates for the cases $H/R=\{0.1;0.08;0.06;0.04\}$, respectively, relative to the time intervals plotted in fig. \ref{timevarbin}. As in Figure \ref{timevarbin}, black line refer to $\dot M_{\rm bin}/\dot M_0$; red line instead is $\dot M_{\rm bin}/M_0$. The frequencies reported are expressed in $\omega/\Omega_{\rm bin}$ with $\omega=2\pi t^{-1}$ and $\Omega_{\rm bin}=2\pi t_{\rm bin }^{-1}$.}\label{periodofig}
\end{figure*}

Figure \ref{timevarbin} shows $\dot M_{\rm sink}$ (red line) and $\dot M_{\rm bin}$ (black line). Starting from the top left panel in Figure \ref{timevarbin} ($H/R=0.1$), we notice that the values of $\dot M_{\rm sink}$ are much smoother than those obtained for $\dot M_{\rm bin}$: the formation of circumprimary and circumsecondary discs (see Figure \ref{colourdensplot}) acts as a buffer for the accretion mechanism, accumulating material and accreting it progressively, smoothing the variability observed instead for the mass flux at $R=a$. This buffering effect was observed by \citet{2014ApJ...783..134F}, who studied the accretion rate in binary systems as a function of the binary mass ratio; we notice also that the qualitative behaviour of the variability of our $H/R=0.1$ case is in very good agreement with their case $q=1$, $H/R=0.1$ (top panel of their Figure 10).

From the other panels in Figure \ref{timevarbin} it can be noticed that the buffering effect progressively disappears as $H/R$ is reduced: in the case $H/R=0.04$ (bottom right panel in Figure \ref{timevarbin}) no circumprimary and circumsecondary discs form, and $\dot M_{\rm sink}$ follows the variability of $\dot M_{\rm bin}$.

Figure \ref{periodofig} shows the frequency analysis of the accretion rates for each $H/R$ plotted in Figure \ref{timevarbin}. Red and black lines show, as previously, $\dot M_{\rm sink}$ and $\dot M_{\rm bin}$, respectively. These periodograms were obtained using Lomb-Scargle analysis on the time interval considered in Figure \ref{timevarbin}. Periodograms of $\dot M_{\rm sink}$ and $\dot M_{\rm bin}$ show that, in thick discs, despite the mass flow $\dot M_{\rm bin}$ is dominated by the frequency $2\Omega_{\rm bin}$, the accretion on to the black holes $\dot M_{\rm sink}$ occurs with lower periodicity, because of the accumulation of the gas into the discs. 

Lowering $H/R$, the absence of the circumprimary and circumsecondary discs causes the power of $\dot M_{\rm sink}$ and $\dot M_{\rm bin}$ to progressively equalize at each frequency, since all the mass that enters in the cavity is accreted by the black holes faster than they are fed from the edge of the cavity. The fundamental frequency of the accretion rate returns to be $\omega=2\Omega_{\rm bin}$ in thin discs. 

It should be finally noticed that the lowest frequency of the periodogram diminishes for low $H/R$: this frequency represents the orbital frequency of the density lump one can observe in Figure \ref{colourdensplot} at the edge of the cavity; the passage of this lump at the pericentre of its orbit generates a boost in the accretion with the same periodicity of the edge of the cavity $\omega_{\rm lump}\sim(GM_{\rm tot}/R_{\rm lump}^3)^{-1/2}$ where $R_{\rm lump}$ is the semi-major axis of the cavity, that results in a peak in the periodogram. Since the cavity becomes larger for low $H/R$, we observe a shift of this peak to lower frequencies for decreasing $H/R$. 

Caution is required in interpreting these results. The variability of the $H/R=0.04$ case is likely affected by the low resolution in the cavity, and the lack of circumprimary and circumsecondary discs may be attributed to the enhanced artificial viscosity in this region. However, some physical considerations also apply. Due to the low accretion rate from the edge of the cavity, circumprimary and circumsecondary discs are expected to become sparser and characterized by lower accretion rates when reducing the disc thickness, implying circumprimary and circumsecondary discs to be fainter than predicted using thicker discs \citep{2015MNRAS.446L..36F} and thus lower luminosities at short wavelengths.

\subsection{The dependence of $\xi$ on viscosity}

Both viscosity and disc temperature are functions of the disc thickness. Performing simulations exclusively for different $H/R$ does not let disentangle whether the observed suppression of the accretion rate is due to viscous or pressure effects. For this reason we performed two additional simulations, S4 and S6 (in Table \ref{tabsim}), with $H/R=\{0.08;0.06\}$ and values of $\alpha$ set in order to obtain the same value of $\nu$ achieved in S3 and S5 with $\alpha=0.1$ and $H/R=\{0.1;0.08\}$, respectively.

The accretion rates of S4 and S6 are consistent with those obtained for S3 and S5, respectively, showing that the accretion rate does not change for the same $\nu$ despite a change in the value of $H/R$. Since $\alpha$ is independent of $H/R$, this suggests that the reduction of the accretion rate for low disc aspect-ratios is due to the reduction of the effective disc viscosity rather than to a variation of the gas pressure. This is consistent with the prediction of \citet{2016MNRAS.tmp..577D} regarding the modifications induced by pressure to the effective gravitational potential: pressure effects allow the gas to overcome gravitationally prohibited regions only for $H/R\gtrsim 0.1$.  

However, the effects of the gas pressure are important for the gas inside the cavity, e.g. for the differential accretion rate \citep{2015MNRAS.452.3085Y,2015MNRAS.447.2907Y}.

\section{Summary and conclusions}\label{concl}

We performed a set of SPH simulations with a circumbinary disc around an equal mass circular binary, varying the disc aspect ratio $H/R$, in order to investigate the dependence of the accretion rate on the disc thickness. 
 
We compared our simulations to a set of reference simulations with a single central object.
We computed the accretion rate at the edge of the cavity $\dot M_{\rm bin}$ in the binary case and compared it to the reference one, $\dot M_0$, from the single central object simulations. Our results for our $H/R=0.1$ case agree with recent literature, in particular they are consistent with \citet{2013MNRAS.436.2997D}, who obtained $\dot M_{\rm bin}/\dot M_0\sim 1$; we also verified the results of \citet{2014ApJ...783..134F} regarding accretion variability and periodicity inside the cavity (although their value $\dot M_{\rm bin}/\dot M_0$ is slightly larger than ours).

While for $H/R\gtrsim 0.1$ the accretion rate in a circular equal mass binary system is $\dot M_{\rm bin}/\dot M_0\sim 1$, for $H/R<0.1$ a linear reduction $\dot M_{\rm bin}/\dot M_0\sim 10H/R$ was observed. These results are summarized in Table \ref{xihor} and Figure \ref{suppraccr}, reporting the values of $\xi(H/R)=\dot M_{\rm bin}/\dot M_0$.

These results have consequences for both the detection and evolution of supermassive black hole binary systems.
If the relationship for the reduction in accretion rate as a function of disc thickness in equation (\ref{linpred}) holds also for the thin discs expected to surround supermassive black holes ($H/R=10^{-2}-10^{-3}$), the accretion rate in supermassive black hole binaries $\dot M_{\rm bin}$ would be reduced of up to a factor $10^{2}$ with respect to the equivalent rate on a single object. 
The low accretion rate implies that these systems are much fainter than normal AGN.

While our simulations do not resolve the individual circumprimary and circumsecondary discs, we do expect that a reduction in $\dot M_{\rm bin}$ implies the formation of lower mass, lower density and lower luminosity minidiscs. This is expected to affect the spectral energy distribution from this kind of systems, since the hottest gas regions may be fainter than so far predicted; in particular reducing the short-wavelength contribution to the continuum spectrum.

Our findings also have important consequences for black hole spin alignment during the merger of a binary black hole system. Firstly, the model of \citet{2015MNRAS.451.3941G} implies that as the accretion rate into the cavity is suppressed, the alignment process of each black hole with its disc is slowed down accordingly. Secondly, with low accretion rates on to the binary the equalization timescale (that is, the timescale required for the two black holes to equalize their mass) is consequently increased, thus justifying a posteriori the assumption of $q=const$, made by \citet{2015MNRAS.451.3941G}. We thus predict that the efficiency of spin alignment during merger is significantly reduced, which will affect the wave form of gravitational waves emitted during the last phases of the merging process and cause a high recoil velocity of the black hole formed after coalescence. 

Low accretion rates on to the black hole binary appears to be important to guarantee the secondary black hole reaches the binary separation at which gravitational waves emissions become dominant for the migration toward the the binary merger \citep{2015MNRAS.452.3085Y}. High accretion rates with a finite mass supply might cause the mass reservoir in the circumbinary disc to be exhausted before the binary reaches the gravitational wave inspiral phase, preventing further angular momentum extraction from the binary and thus implying the stallation of the migration process. 
In this context the pileup of material at the cavity edge due to the suppression of accretion provides a stronger binary-disc coupling and, as a consequence, a more effective delivery of angular momentum to the gaseous disc \citep{2013ApJ...774..144R}, shortening the migration time.

We finally emphasize that our results are essentially scale-free and can thus be extended to all black hole binary systems, providing predictions of accretion-dynamics in any mass regime. Equal mass, circular binaries with stellar mass black holes (such as those simulated here) are of particular interest due to the recent discovery of the gravitational wave source GW150914 \citep{2016PhRvL.116f1102A}. Some suggestion, although uncertain, for the occurrence of an electromagnetic counterpart to the gravitational wave emission in this particular source has also been made (e.g., \citealp{2016arXiv160203920C}). Such electromagnetic counterparts are most naturally determined by gas accretion prior, during and after the merger. Our results emphasize the important role that the disc thickness has in determining the tidal torques ability to act as a dam for the gas, which on the one hand might prevent gas flow from the circumbinary disc onto the binary, as discussed here, while on the other hand it might prevent gas flow from the individual disc to the circumbinary environment, as the gas is squeezed during the gravitational wave driven inspiral \citep{2016MNRAS.457..939C}.

Possible improvements to this work consist primarily of extending the parameter space under investigation, in particular considering the case of binary with eccentric orbits and exploring non-unitary mass ratios. Different mass-ratios would provide new data for the differential accretion rates of the individual black hole, on which the model for spin alignment of \citet{2015MNRAS.451.3941G} is based. 

Secondly, a broader range of $H/R$ values should be explored. However, while exploring the $H/R>0.13$ regimes would be straightforward, studying aspect ratios $H/R<0.02$ would be more challenging since both spatial and accretion resolution were not well achieved already in our $H/R=0.02$ case. A possible solution could be to restrict the area of the study to the region of the cavity providing a constant mass flux from outer radii as done for differential accretion rate studies \citet{1997MNRAS.285...33B, 2005ApJ...623..922O, 2010ApJ...708..485H, 2015MNRAS.447.2907Y}. Finally, while here we discuss only the case of discs that are aligned with the orbital plane of the binary, it would also be interesting to study the problem of misaligned circumbinary discs \citep{2015ApJ...800...96L}.

\section*{Acknowledgements}
We thank Cathie Clarke for useful discussions. We thank Davide Gerosa for a careful reading of the manuscript and precious comments.
We acknowledge the referee for constructive comments and useful suggestions that improved the manuscript.
ER and GL acknowledge financial support from PRIN MIUR 2010-2011, project ``The Chemical and Dynamical Evolution of the Milky Way and Local Group Galaxies'', prot. 2010LY5N2T. 
DJP is supported by a Future Fellowship from the Australian Research Council (FT130100034).
Some plots in this paper were produced using \textsc{splash} \citep{2007PASA...24..159P}.




\bibliography{./biblio}


\appendix

\section{Code Units}\label{codeunitsec}

Simulations are completely scalable. We have chosen parameters to allow the easy conversion to physical units. 
Our setup has indeed $M_{\rm tot}=1$, $a=1$ and $G=1$; this gives $t_{\rm bin}=2\pi$, where $t_{\rm bin}=2\pi (GM/a^3)^{-1/2}$. 
Once the physical total mass of the binary $M_{\rm tot}$ and the binary distance $a$ have been fixed, lengths $l_{\rm code}$, time $t_{\rm code}$, velocity $v_{\rm code}$, density $\rho_{\rm code}$, surface density $\Sigma_{\rm code}$ and accretion rate $\dot M_{\rm code}$ in code units have the following rescaling to physical units:
\begin{align}
l_{\rm phys}&= 1.496\times 10^{13}\, a_{\rm AU}\cdot l_{\rm code} \;{\rm cm}=3.085\times 10^{\rm 18}\, a_{\rm pc} \cdot l_{\rm code}\;{\rm cm},	
\end{align}
\begin{align}
t_{\rm phys}&=\frac{1}{2\pi}\sqrt{\frac{a^3_{\rm AU}}{M_{{\rm tot},M_\odot}}}\cdot t_{\rm code}\; {\rm yr},\\
&= 1.487\times 10^{3}\sqrt{\frac{a^3_{\rm pc}}{M_{{\rm tot},10^8\cdot M_\odot}}} \cdot t_{\rm code} \; {\rm yr},
\end{align}
\begin{align}
v_{\rm phys}&= 29.78 \sqrt{\frac{M_{{\rm tot},M_\odot}}{a_{\rm AU}}}\cdot v_{\rm code}\; {\rm km\, s^{-1}}\\
&=6.561\times 10^2\sqrt{\frac{ M_{{\rm tot},10^8\cdot M_\odot}}{a_{\rm pc}}}\cdot v_{\rm code}\; {\rm km\, s^{-1}},
\end{align}
\begin{align}
\rho_{\rm phys}&=5.941\times 10^{-6} \frac{M_{{\rm tot},M_\odot}}{a^3_{\rm AU}}\cdot\rho_{\rm code}\; {\rm g\,cm^{-3}}\\
& =6.435\times 10^{-14}\frac{M_{{\rm tot},10^8\cdot M_\odot}}{a_{\rm pc}^3}\cdot\rho_{\rm code}\; {\rm g\,cm^{-3}},
\end{align}
\begin{align}
\Sigma_{\rm phys}&=8.887\times 10^{7}\frac{M_{{\rm tot}, M_\odot}}{a_{\rm AU}^2}\cdot\Sigma_{\rm code}\; {\rm g\,cm^{-2}}\\
&=1.969\times 10^{5}\frac{M_{{\rm tot},10^8\cdot M_\odot}}{a_{\rm pc}^2}\cdot\Sigma_{\rm code}\; {\rm g\,cm^{-2}},
\end{align}
\begin{align}
\dot M_{\rm phys}&=2\pi\cdot\sqrt{\frac{M^3_{{\rm tot},\odot}}{a^3_{\rm AU}}}\cdot \dot M_{\rm code}\; {\rm M_{\odot}\,yr^{-1}},\\
&=6.724\times 10^4\sqrt{\frac{M^{3}_{{\rm tot},10^8\cdot M_\odot}}{a^3_{\rm pc}}}\cdot \dot M_{\rm code}\; {\rm M_{\odot}\,yr^{-1}},
\end{align}
where $l_{\rm phys}$ is the physical length, $t_{\rm phys}$ is the physical time, $v_{\rm phys}$ is the physical speed, $\rho_{\rm phys}$ the physical density, $\Sigma_{\rm phys}$ the physical surface densisty of the disc and  $M_{\rm phys}$ the physical accretion rate; $a_{\rm AU}$ and $a_{\rm pc}$ are the binary separation in ${\rm AU}$ and ${\rm pc}$ units, respectively; $M_{{\rm tot}, M_\odot}$ and $M_{{\rm tot},10^8\cdot M_\odot}$ are the total mass of the binary in solar masses, $M_\odot$, and $10^8 M_\odot$ units, respectively.


\bsp	
\label{lastpage}
\end{document}